\documentclass[twocolumn, twocolappendix]{aastex63}
\pdfoutput=1

\received{2024 April 2}
\revised{2024 July 3}
\accepted{2024 July 13}
\submitjournal{ApJ}

\shorttitle{Radio scrutiny of X-ray-weak low-mass AGNs}
\shortauthors{Paul et al.}

\graphicspath{{./}}
\usepackage{multirow}
\usepackage{amsmath}

\defcitealias{GH07s}{G\&H07}

\newcommand{\lrlx}{L_{\rm R}/L_{\rm X}}
\newcommand{\Rx}{\mathcal{R}_{\rm X}}
\newcommand{\Rxexp}{\mathcal{R}_{\rm X, exp}}
\newcommand{\Ro}{\mathcal{R}_{\rm O}}
\newcommand{\ha}{H$\alpha$}
\newcommand{\aox}{\alpha_{\rm ox}}
\newcommand{\daox}{\Delta \alpha_{\rm ox}}
\newcommand{\fweak}{f_{\rm weak}}

\newcommand{\alphar}{\alpha_{\rm r}}
\newcommand{\ledd}{\ell_{\rm Edd}}

\begin{document}

\title{Radio Scrutiny of the X-ray-Weak Tail of Low-Mass Active Galactic Nuclei: \\
A Novel Signature of High-Eddington Accretion?}

\correspondingauthor{Jeremiah D. Paul} 
\email{jeremiahp@unr.edu}

\author[0000-0003-0040-3910]{Jeremiah D. Paul}
\affiliation{Department of Physics, University of Nevada, Reno, NV 89557, USA}

\author[0000-0002-7092-0326]{Richard M. Plotkin}
\affiliation{Department of Physics, University of Nevada, Reno, NV 89557, USA}
\affiliation{Nevada Center for Astrophysics, University of Nevada, Las Vegas, NV 89154, USA}

\author[0000-0002-0167-2453]{W. N. Brandt}
\affiliation{Department of Astronomy \& Astrophysics, 525 Davey Lab, The Pennsylvania State University, University Park, PA 16802, USA}
\affiliation{Institute for Gravitation and the Cosmos, The Pennsylvania State University, University Park, PA 16802, USA}
\affiliation{Department of Physics, 104 Davey Lab, The Pennsylvania State University, University Park, PA 16802, USA}

\author{Christopher H. Ellis}
\affiliation{Department of Physics, University of Nevada, Reno, NV 89557, USA}

\author[0000-0001-5802-6041]{Elena Gallo}
\affiliation{Department of Astronomy, University of Michigan, 1085 S. University Ave, Ann Arbor, MI 48109, USA}

\author[0000-0002-5612-3427]{Jenny E. Greene}
\affiliation{Department of Astrophysical Sciences, 4 Ivy Lane, Princeton University, Princeton, NJ 08544, USA}

\author[0000-0001-6947-5846]{Luis C. Ho}
\affiliation{Kavli Institute for Astronomy and Astrophysics, Peking University, Beijing 100871, China}
\affiliation{Department of Astronomy, School of Physics, Peking University, Beijing 100871, China}

\author[0000-0001-9324-6787]{Amy E. Kimball}
\affiliation{National Radio Astronomy Observatory, 1011 Lopezville Road, Socorro, NM 87801, USA}

\author[0000-0001-6803-2138]{Daryl Haggard}
\affiliation{Department of Physics, McGill University, 3600 rue University, Montréal, Québec, H3A 2T8, Canada}
\affiliation{Trottier Space Institute, 3550 Rue University, Montréal, Québec, H3A 2A7, Canada}

\begin{abstract}
The supermassive black holes ($M_{\rm BH} \sim 10^{6}$--$10^{10}~M_\odot$) that power luminous active galactic nuclei (AGNs), i.e., quasars, generally show a correlation between thermal disk emission in the ultraviolet (UV) and coronal emission in hard X-rays. In contrast, some ``massive" black holes (mBHs; $M_{\rm BH} \sim 10^{5}$--$10^{6}~M_\odot$) in low-mass galaxies present curious X-ray properties with coronal radiative output up to 100$\times$ weaker than expected. To examine this issue, we present a pilot study incorporating Very Large Array radio observations of a sample of 18 high-accretion-rate (Eddington ratios $L_{\rm bol}/L_{\rm Edd} > 0.1$), mBH-powered AGNs ($M_{\rm BH} \sim 10^{6}~M_\odot$) with Chandra X-ray coverage. Empirical correlations previously revealed in samples of radio-quiet, high-Eddington AGNs indicate that the radio--X-ray luminosity ratio, $L_{\rm R}/L_{\rm X}$, is approximately constant. Through multiwavelength analysis, we instead find that the X-ray-weaker mBHs in our sample tend toward larger values of $L_{\rm R}/L_{\rm X}$ even though they remain radio-quiet per their optical--UV properties. This trend results in a tentative but highly intriguing correlation between $L_{\rm R}/L_{\rm X}$ and X-ray weakness, which we argue is consistent with a scenario in which X-rays may be preferentially obscured from our line of sight by a ``slim" accretion disk. We compare this observation to weak emission-line quasars (AGNs with exceptionally weak broad-line emission and a significant X-ray-weak fraction) and conclude by suggesting that our results may offer a new observational signature for finding high-accretion-rate AGNs.
\end{abstract}


\section{Introduction} \label{sec:intro}

In the last $\approx$\,20~yr the number of discovered low-mass active galactic nuclei (AGNs) powered by black holes with $M_{\rm BH} \lesssim 10^{6}~M_\odot$ has grown from only two secure candidates \citep{Filippenko03, Barth04} to homogeneously selected samples reaching hundreds of objects \citep[e.g.,][]{Greene04, GH07s, Barth08, Dong12b, Kamizasa12, Reines13, Lemons15, Liu18, Chilingarian18, Martinez20, Shin22, Salehirad22, Zou23}. The low-mass AGNs identified in many of the above samples are in the so-called ``massive" black hole range (mBH; $M_{\rm BH} \sim 10^{5}$--$10^{6}~M_\odot$), straddling the boundary between supermassive black holes (SMBHs; $M_{\rm BH} \sim 10^{6}$--$10^{10}~M_\odot$) and intermediate-mass black holes ($M_{\rm BH} \sim 10^{2}$--$10^{5}~M_\odot$). Due to several observational biases and selection effects, the majority of high-confidence, mBH-powered AGNs tend to be radio-quiet\footnote{We adopt the definition whereby radio-quiet AGNs have radio-to-optical flux density ratios $\Ro=f\textsubscript{5 GHz}/f\textsubscript{4400 {\AA}} < 10$, where $f\textsubscript{\rm 5 GHz}$ and $f\textsubscript{4400 {\AA}}$ are the radio and optical flux densities at 5~GHz and 4400~{\AA}, respectively \citep[e.g.,][]{Kellermann89,Stocke92}.} \citep[e.g.,][]{Greene06} and have moderate to high accretion rates (Eddington ratios $\ledd \gtrsim 0.01$)\footnote{We denote the Eddington ratio via $\ledd = L_{\rm bol}/L_{\rm Edd}$, where $L_{\rm bol}$ is the bolometric luminosity and $L_{\rm Edd} = 1.26\times 10^{38}(M/M_\odot)$~erg\,s$^{-1}$ is the Eddington luminosity.}

With such properties, mBH AGNs offer an intriguing opportunity to begin extending our present understanding of black-hole accretion in relatively well-studied, Type 1 (unobscured) quasar populations toward the intermediate-mass regime. In turn, lower-mass black holes will help advance our understanding of numerous topics, including how SMBHs form and grow (e.g., \citealt{Inayoshi20} and references therein; \citealt{Wang21, Farina22, Fan23, Larson23, Maiolino23, Lai24}) and how black-hole mass impacts accretion/jet physics \citep[e.g.,][]{Falcke95, Gultekin14, Gultekin22} and host-galaxy feedback \citep[e.g.,][]{Fabian12, Kormendy13, King15}. Unfortunately, there are still significant obstacles to finding and securely confirming extremely low-mass candidates (at present, $M_{\rm BH} \lesssim 10^{5}~M_\odot$; for recent comprehensive reviews, see, e.g., \citealt{Reines16, Greene20, Reines22}). A better understanding of the physics and observational signatures of accretion in the mBH and intermediate-mass regimes is needed.

Na{\"i}vely, we expect mBH AGNs to display a relationship between disk and coronal emission similar to that observed for Type 1 quasars, which generally follow an anticorrelation between their X-ray--ultraviolet (UV) broadband spectral indices and their UV monochromatic luminosities \citep[e.g.,][]{Steffen06, Just07, Lusso10, Chiaraluce18, Timlin20, Timlin21, Maithil24}. In fact, mBH AGNs display a higher average ratio of X-ray to UV luminosities than Type 1 quasars \citep[e.g.,][]{Desroches09, Dong12a, Liu18}. This observation (combined with the expectation that black holes of lower mass should possess hotter accretion disks; e.g., \citealt{Shakura73}) has been taken to suggest that, compared to quasars, mBHs have hotter thermal disk emission that peaks in the soft X-ray waveband ($\sim$\,0.1~keV), and this is combined with a hard X-ray ($\gtrsim$\,1--2~keV) contribution from a less efficiently cooled corona \citep{Haardt93, Done12, Dong12a}. Curiously, however, mBH AGNs also show an unexpectedly wide dispersion in their X-ray to UV luminosity ratios, with a perplexingly long X-ray-weak tail \citep[e.g.,][]{Dong12a, Liu18} where the coronal radiative output can appear up to 100$\times$ weaker than expected. 

The cause of this X-ray-weak tail remains unclear. The two leading explanations proposed are: (1) some mBH AGNs are intrinsically X-ray weak \citep{Dong12a}; and (2) some mBH AGNs experience absorption or scattering of X-rays despite remaining unobscured in the optical--UV \citep{Desroches09, Plotkin16}. In case (1), intrinsic X-ray weakness could be caused by less efficient coronal upscattering of disk photons into hard X-rays, from which we may expect a softer (i.e., steeper) X-ray spectrum (e.g., \citealt{Leighly07b, Leighly07a, Nardini19, Zappacosta20, Laurenti22}; however, see also \citealt{Liu21, Wang22}). In case (2), preferential attenuation of observed X-rays requires a medium that shields only the X-ray emitting region but not the optical--UV \citep[e.g.,][]{Wu11, Luo15, Ni18, Ni22}, and X-ray-weak objects should display a comparatively harder (i.e., flatter) X-ray spectrum. 

Building on the latter case, such shielding could be achieved by an advection-dominated, geometrically thick, ``slim" accretion disk \citep[e.g.,][]{Abramowicz88,Czerny19} and its related wide-angle outflows \citep[e.g.,][]{Murray95, Castello-Mor17, Giustini19, Jiang19, Naddaf22}. In this scenario, the innermost regions of the disk are inflated by radiation pressure. X-ray weakness would then be observed as a byproduct of orientation (allowing unification of the weak and normal populations), as larger inclination angles leave X-rays increasingly obscured along our line of sight in a manner similar to that proposed for weak emission-line quasars (a population of Type 1 quasars with exceptionally weak broad-line emission and a substantial X-ray-weak fraction; e.g., \citealt{Luo15, Ni18, Ni22}). Unfortunately, X-ray spectral studies on mBH AGN samples have yet to distinguish conclusively between the two scenarios, and while a stacking analysis of the X-ray-weakest objects does suggest flatter X-ray spectra that could be caused by scattering/absorption, uncertainties are extremely large (\citealt{Plotkin16}). 

A multiwavelength approach incorporating observations of nuclear radio emission from radio-quiet AGNs may help distinguish between the weak-corona and slim-disk scenarios, thereby elucidating the source of the mBH X-ray-weak tail. The dominant source of radio emission from radio-quiet AGNs is still under investigation and may involve a number of physical mechanisms, including disk--corona activity, weak jet bases, free-free scattering in photoionized circumnuclear gas, star formation activity, and/or poorly collimated, sub-relativistic outflows such as winds (see \citealt{Panessa19} for a review). Nevertheless, clear empirical correlations established between the hard X-ray and GHz radio luminosities ($L_{\rm X}$ and $L_{\rm R}$, respectively) of radio-quiet AGNs show the ratio of luminosities ($\lrlx$) to be approximately constant at a given physical radio emission scale \citep[e.g.,][]{Panessa07, Panessa15, Panessa22, Laor08, Behar15, Yang20}, with higher spatial resolution observations of core radio emission generally showing lower values of $\lrlx$ as extended radio emission is resolved out \citep[e.g.,][]{Chen23, Shuvo23, Wang23}.\footnote{Note that while this may imply a causal connection between X-ray and radio emission mechanisms, such a link remains under investigation \citep[e.g.,][]{Behar20, Chen22}.} While a similar relationship has been observed for some mBH AGNs \citep[][]{Gultekin14, Gultekin22}, no prior radio study has sampled the full range of X-ray weakness displayed by mBH AGNs. 

Herein, we describe a multiwavelength pilot study to test the origin of X-ray weakness in mBHs. We do this using a sample of 18 high-Eddington mBH AGNs from the catalog identified by \citet[][hereafter \citetalias{GH07s}]{GH07s}, combining new and archival Karl G. Jansky Very Large Array (VLA) radio observations with archival Chandra X-ray Observatory data covering the full observed range of X-ray weakness. Crucially, even the smallest-scale nuclear radio emission that could be emitted, e.g., from a compact radio corona or jet base, is likely to originate outside the primary X-ray shielding zone and thus remain unshielded: \citet{Laor08} show that the minimum radius of a coronal synchrotron sphere producing GHz radio emission is expected to be $>$\,100$\times$ the radius of the X-ray-emitting core (see their Eq.~19 and related discussion). As we will show, if X-ray weakness in high-Eddington mBHs is caused predominantly by shielding instead of weak coronae, then we are likely to see $\lrlx$ correlate with X-ray weakness, deviating from the empirical $\lrlx \sim$~constant relationship. The paper is organized as follows: In Section \ref{sec:obs} we describe our observations along with our data reduction and analysis methods. We present our results in Section \ref{sec:results} and discuss their implications in Section \ref{sec:discussion}. Finally, we summarize our conclusions in Section \ref{sec:summ}. We relegate most per-object discussion or caveats to the Appendix. We adopt the cosmological parameters $H_0=70$ km s$^{-1}$ Mpc$^{-1}$, $\Omega_{\rm M}=0.3$, and $\Omega_{\rm \Lambda}=0.7$. All uncertainties are reported at the $\pm$1$\sigma$ confidence level unless otherwise noted.

\section{Observations and Data Reduction} \label{sec:obs}

\subsection{Sample Selection} \label{subsec:samp}

The \citetalias{GH07s} catalog consists of 174 high-confidence, unobscured low-mass AGNs identified based on the presence of broad H$\alpha$ emission in optical spectra from the Sloan Digital Sky Survey (SDSS; \citealt{SDSS1}). These AGNs are predominantly (99\%) radio-quiet, tend toward high accretion rates (average Eddington ratio $\langle \ledd \rangle \approx 0.3$; median value $\ledd = 0.4$), and are in the mBH range (average mass $\langle M_{\rm BH} \rangle \approx 1.2 \times 10^{6}~M_\odot$; median value $M_{\rm BH} = 1.3 \times 10^{6}~M_\odot$). Of these objects, 65 have been observed in the X-ray by Chandra (studies performed by \citealt{GH07b, Desroches09, Dong12a, Gultekin14}). 

Our sample of 18 targets was drawn from the 65 high-confidence \citetalias{GH07s} AGNs with Chandra X-ray coverage, which we limited to 54 objects with $\ledd > 0.1$. Note that while we adopt the values of $\ledd$ reported by \citetalias{GH07s}, \citet{Plotkin16} explore multiple methods for applying bolometric luminosity corrections and suggest that $\ledd$ from \citetalias{GH07s} may be systematically underestimated by up to an order of magnitude. This has recently been corroborated by \citet{Cho23}, who showed that \citetalias{GH07s} mBH masses may be overestimated by up to 0.7 dex, as well as through general conclusions that typical broad line region size--luminosity relationships tend to overestimate $M_{\rm BH}$ in high-Eddington, low-luminosity (low-mass) AGNs \citep[e.g.,][]{Du18, Maithil22, Woo24}. We note this to stress the likelihood that our sample is accreting at near- or super-Eddington rates above $\ledd = 0.3$.

\begin{figure*}[t]
\gridline{\fig{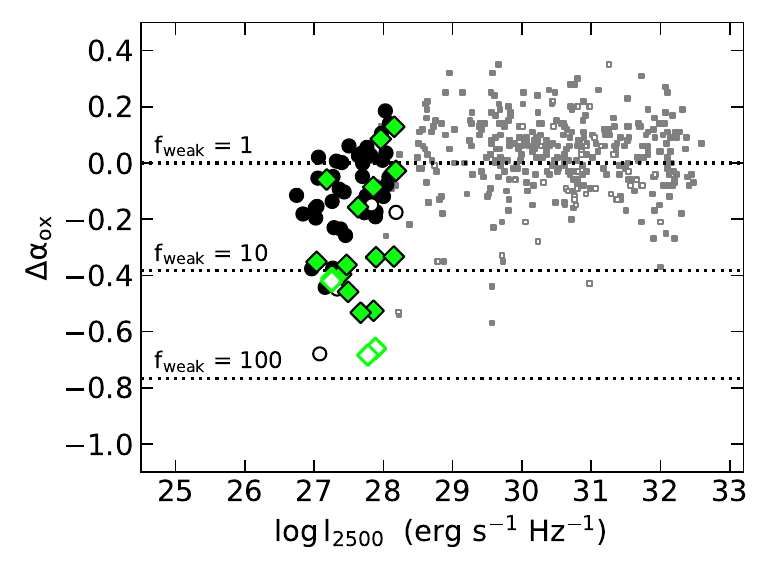}{0.49\textwidth}{\textbf{(a)}}
          \fig{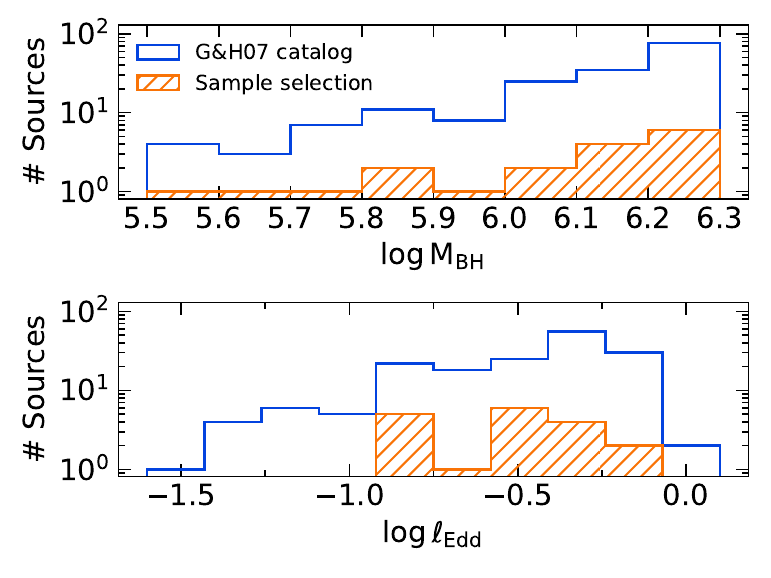}{0.49\textwidth}{\textbf{(b)}}}
\caption{\textbf{(a):} X-ray weakness (represented by $\daox$; more negative values are X-ray weaker) vs.\ 2500~{\AA} monochromatic luminosity. Horizontal dotted lines denote $\fweak$ at values of 1, 10, and 100 (see Section \ref{subsec:samp}). Note that $\fweak = 1$ ($\daox = 0$) marks the \citet{Just07} best-fit $\aox$--$l_{\rm 2500}$ relationship. Our target sample of 18 \citetalias{GH07s} low-mass AGNs is shown by green diamonds ($\daox$ values are from our recalculations; see Section \ref{subsec:Xobs}), while the remainder of the \citetalias{GH07s} catalog is represented by black circles ($\daox$ values from \citealt{Desroches09} and \citealt{Dong12a}). Filled and empty symbols denote X-ray detections and non-detections, respectively. Grey squares represent the SDSS quasar samples used by \citet{Steffen06} and \citet{Just07} to define $\daox$. Our sample spans the full range of X-ray weaknesses observed in the \citetalias{GH07s} catalog.
\textbf{(b):} Histograms showing the $M_{\rm BH}$ (top) and $\ledd$ (bottom) distributions of our sample selection (orange hatched region) compared to the \citetalias{GH07s} parent sample (blue outlined region).
\label{fig:weak_tail}}
\end{figure*}

Of the 54 objects satisfying the above criteria, 7 had already been observed by VLA and examined in radio and X-ray by \citet{Gultekin14}; we designated these 7 AGNs as the ``archival" portion of our sample. We found this archival sample to be predominantly X-ray normal (a result of the selection criteria used by \citealt{Gultekin14}) per the following description. To define X-ray weakness, we parameterized the ratio of X-ray to UV specific luminosities following standard convention, beginning with the broadband spectral index, 
\begin{equation} \label{eq1}
\aox = 0.38\log(l_{\rm 2 keV} / l_{\rm 2500}),
\end{equation}
where $l_{\rm 2 keV}$ and $l_{\rm 2500}$ are the unabsorbed monochromatic luminosities at rest-frame 2~keV and 2500~{\AA}, respectively \citep[][]{Tananbaum79}.\footnote{As described in Section \ref{subsec:Xobs}, $l_{\rm 2500}$ is derived using a method designed to mitigate the risk of host-galaxy contamination.} We also adopted the customary X-ray deviation parameter,
\begin{equation} \label{eq2}
\daox = \aox - \alpha_{\rm ox,qso},
\end{equation}
where $\alpha_{\rm ox,qso}$ is the value predicted by the $\aox$--$l_{\rm 2500}$ relationship displayed by broad-line quasars.\footnote{For consistency with prior works, we adopt the best-fit $\alpha_{\rm ox,qso}$ relationship given by Eq.~(3) of \citet{Just07}. \citet{Timlin20} suggest an intrinsic scatter of $\pm 0.11$~dex.} Finally, we adopted the X-ray weakness factor,
\begin{equation} \label{eq3}
\fweak = 10^{- \daox / 0.38},
\end{equation}
so that X-ray-weaker objects have larger $\fweak$ factors. As seen in Figure \ref{fig:weak_tail}(a), some mBH AGNs could be up to 100$\times$ weaker in X-ray than expected from the established quasar relationship (see also \citealt{Dong12a}; \citealt{Plotkin16}). The majority of the archival sample is X-ray normal with $\fweak < 6$ (i.e., $\daox > -0.3$). 

To examine the full range of X-ray \textit{weak} low-mass AGNs, from the 47 remaining objects we finally selected those with $\fweak > 6$ (based on values of $\daox < -0.3$ taken directly from \citealt{Desroches09} and \citealt{Dong12a}). This restriction left 11 targets for new VLA radio observations (discussed in Section \ref{subsec:radio_obs}); we designated these AGNs as the ``new" portion of our sample. Two-sample Kolmogorov–Smirnov (K-S) tests between our combined (archival + new) 18-object selection and the parent remainder show that their optically derived parameters follow similar distributions, including $l_{\rm 2500}$ ($p = 0.6$), $M_{\rm BH}$ ($p = 0.9$; see Figure \ref{fig:weak_tail}(b)), and redshift ($p = 0.5$; all objects are at $z \lesssim 0.15$, although we note that no explicit redshift cut was applied in our sample selection). The optical--X-ray properties of the 18 objects in our full sample are given in Table \ref{tab:sample}. Note that we do not tabulate the literature values for $l_{\rm 2 keV}$, $\aox$, $\daox$, or $\fweak$, but we instead give updated values derived from re-reduction of the archival X-ray observations, as described below. Throughout this paper, we refer to individual targets using their ``GH ID" designation from \citetalias{GH07s}.

\begin{deluxetable*}{lcccccccccc}[t]
\tablenum{1}
\tablecaption{Sample Optical and X-ray Properties\label{tab:sample}}
\tablehead{
\colhead{Source Name}  &  \colhead{GH ID}  &  \colhead{$z$}  &  \colhead{$\log M_{\rm BH}$}  &  \colhead{$\ell_{\rm Edd}$}  &  \colhead{$\log l_{\rm 2500}$}  & \colhead{$\log l_{\rm 2keV}$}  &  \colhead{$\aox$}  &  \colhead{$\daox$}  &  \colhead{$\fweak$}  &  \colhead{Ref.} \\
\colhead{(SDSS J)}  &  \colhead{}  &  \colhead{}  &  \colhead{($M_\odot$)}  &  \colhead{}  &  \colhead{(erg~s$^{-1}$~Hz$^{-1}$)}  &  \colhead{(erg~s$^{-1}$~Hz$^{-1}$)}  &  \colhead{}  &  \colhead{}  &  \colhead{}  &  \colhead{} \\
\colhead{(1)}  &  \colhead{(2)}  &  \colhead{(3)}  &  \colhead{(4)}  &  \colhead{(5)}  &  \colhead{(6)}  &  \colhead{(7)}  &  \colhead{(8)}  &  \colhead{(9)}  &  \colhead{(10)}  &  \colhead{(11)}  
}
\startdata
\multicolumn{11}{c}{New VLA sample}\\
\hline
032515.58$+$003408.4  & 25 & 0.102 & 6.2 & 0.32 & 27.86 & 23.37 & $-$1.72 & $-$0.53 & 23.46 & 1 \\
092438.88$+$560746.8  & 73 & 0.025 & 6.1 & 0.16 & 27.49 & 23.32 & $-$1.60 & $-$0.46 & 15.65 & 2 \\
094310.11$+$604559.1  & 80 & 0.074 & 6.0 & 0.50 & 27.89 & $<$23.04 & $< -$1.86 & $< -$0.66 & $>$52.47 & 3 \\
105755.66$+$482501.9  & 104 & 0.073 & 5.8 & 0.16 & 27.26 & $<$23.29 & $< -$1.52 & $< -$0.41 & $>$11.79 & 2 \\
131651.29$+$055646.9  & 157 & 0.055 & 6.3 & 0.16 & 27.63 & 24.19 & $-$1.32 & $-$0.16 & 2.57 & 2 \\
131926.52$+$105610.9  & 160 & 0.064 & 5.9 & 0.40 & 27.67 & 23.24 & $-$1.70 & $-$0.53 & 24.45 & 2 \\
144052.60$-$023506.2  & 185 & 0.044 & 6.1 & 0.32 & 27.78 & $<$22.91 & $< -$1.87 & $< -$0.68 & $>$60.49 & 2 \\
162636.40$+$350242.1  & 211 & 0.034 & 5.7 & 0.32 & 27.39 & 23.42 & $-$1.53 & $-$0.40 & 10.76 & 2 \\
163159.59$+$243740.2  & 213 & 0.043 & 5.8 & 0.32 & 27.47 & 23.55 & $-$1.50 & $-$0.36 & 8.80 & 2 \\
163228.89$-$002843.9  & 214 & 0.070 & 6.0 & 0.13 & 27.25 & $<$23.27 & $< -$1.53 & $< -$0.42 & $>$12.44 & 2 \\
165636.98$+$371439.5  & 215 & 0.063 & 6.3 & 0.25 & 27.89 & 23.89 & $-$1.54 & $-$0.34 & 7.48 & 2 \\
\hline
\multicolumn{11}{c}{Archival VLA sample}\\
\hline
082443.28$+$295923.5  & 47 & 0.025 & 5.6 & 0.16 & 27.03 & 23.30 & $-$1.43 & $-$0.35 & 8.27 & 2 \\
091449.05$+$085321.1  & 69 & 0.140 & 6.2 & 0.63 & 28.16 & 25.27 & $-$1.11 & 0.13 & 0.46 & 4 \\
101246.59$+$061604.7  & 87 & 0.078 & 6.1 & 0.32 & 27.85 & 24.52 & $-$1.28 & $-$0.09 & 1.67 & 4 \\
110501.98$+$594103.5  & 106 & 0.034 & 5.5 & 0.32 & 27.18 & 24.16 & $-$1.16 & $-$0.06 & 1.42 & 2 \\
124035.82$-$002919.4  & 146 & 0.081 & 6.3 & 0.50 & 28.15 & 24.06 & $-$1.57 & $-$0.33 & 7.38 & 3 \\
140829.27$+$562823.4  & 174 & 0.134 & 6.1 & 0.50 & 27.96 & 25.03 & $-$1.12 & 0.09 & 0.60 & 4 \\
155909.62$+$350147.5  & 203 & 0.031 & 6.2 & 0.63 & 28.18 & 24.88 & $-$1.27 & $-$0.03 & 1.18 & 2\\
\enddata
\tablecomments{Column (1): object SDSS name. Column (2): \citetalias{GH07s} ID. Column (3): spectroscopic redshift adopted from SDSS DR12 \citep{Alam15}. Column (4): logarithm of black hole mass estimate from \citetalias{GH07s}. Column (5): Eddington ratio $\ledd = L_{\rm bol}/L_{\rm Edd}$ from \citetalias{GH07s}. Column (6): logarithm of UV monochromatic luminosity at 2500~{\AA} (see Section \ref{subsec:Xobs}). Column (7): logarithm of 2~keV monochromatic luminosity (see Section \ref{subsec:Xobs}). Columns (8--10): X-ray--UV broadband spectral index, X-ray deviation parameter, and X-ray weakness factor calculated from re-reduced archival observations (see Sections \ref{subsec:samp}, \ref{subsec:Xobs}). Column (11): reference for original X-ray observation; (1) \citet{Desroches09}, (2) \citet{Dong12a}, (3) \citet{GH07b}, and (4) \citet{Gultekin14}.
}
\end{deluxetable*}

\subsection{Archival X-ray Observations} \label{subsec:Xobs}

To ensure a uniform analysis, we downloaded and re-reduced all archival Chandra X-ray observations for our full 18-object sample. Our X-ray results are reported in Table \ref{tab:xray}. Data were reduced following standard procedures with the Chandra Interactive Analysis of Observations (CIAO) software v14.4, with CALDB v4.9.8 \citep{Fruscione06}. We first re-processed the data with {\tt chandra\_repro} and confirmed that there were no periods of background flaring during our observations. Next we ran {\tt wavdetect} on each observation to search for X-ray sources in each galaxy (setting the wavelet {\tt scales} parameter to ``1 2 4 6 8 12 16 24 32'' and {\tt sigthresh} to 10$^{-6}$). X-ray sources were found in the nuclei of 14 galaxies, with a wide range of counts reported by {\tt wavdetect}, 6--1400 over 0.5--7.0~keV. For these 14 X-ray sources, we measured net full-band (0.5--7.0~keV) count rates using {\tt srcflux}, adopting the default source aperture to enclose 90\% of the point-spread-function (psf) at 1 keV (aperture corrections were performed in {\tt srcflux} using {\tt psfmethod=arfcorr}). For the four non-detections, after confirming from visual inspection that no X-ray source was present, we measured the local background near the nucleus of each galaxy. We then placed upper limits on the net count rate at the 99\% confidence level, assuming Poisson statistics in the presence of background \citep{Kraft91}.

\begin{deluxetable*}{lcccccccc}[t]
\tablenum{2}
\tablecaption{Archival X-ray Observations\label{tab:xray}}
\tablehead{
\colhead{GH ID}  &  \colhead{ObsID}  &  \colhead{$\tau_{\rm exp}$}  &  \colhead{$N_{\rm full}$}  &  \colhead{$R_{\rm full}$}  &  \colhead{$\log N_{\rm H,fit}$}  &  \colhead{$\Gamma$}  &  \colhead{$\log f_{\rm 2-10keV}$}  &  \colhead{C-stat/dof}  \\
\colhead{}  &  \colhead{}  &  \colhead{(ks)}  &  \colhead{(counts)}  &  \colhead{(counts~ks$^{-1}$)}  &  \colhead{(cm$^{-2}$)}  &  \colhead{}  &  \colhead{(erg~s$^{-1}$~cm$^{-2}$)}  &  \colhead{}  \\
\colhead{(1)}  &  \colhead{(2)}  &  \colhead{(3)}  &  \colhead{(4)}  &  \colhead{(5)}  &  \colhead{(6)}  &  \colhead{(7)}  &  \colhead{(8)}  &  \colhead{(9)}
}
\startdata
\multicolumn{9}{c}{New VLA sample}\\
\hline
25 & 7733 & 4.52 & $8.7 \pm 3.2$ & $1.9 \pm 0.7$ & 20.9\textsuperscript{*} & \nodata & $-14.08 \pm 0.16$ & \nodata \\
73 & 11449 & 1.94 & $48.7 \pm 7.4$ & $9.1 \pm 2.4$ & $<21.2$ & $1.23 \pm 0.20$ & $-12.64 \pm 0.11$ & 38.7/44 \\
80 & 5661 & 5.03 & $<8.3$ & $<1.7$ & 20.4\textsuperscript{*} & \nodata & $<-14.13$ & \nodata \\
104 & 11455 & 1.97 & $<6.3$ & $<3.2$ & 20.2\textsuperscript{*} & \nodata & $<-13.87$ & \nodata \\
157 & 11469 & 1.97 & $34.0 \pm 6.2$ & $9.8 \pm 2.4$ & $22.1 \pm 0.3$ & $1.38 \pm 0.33$ & $-12.52 \pm 0.14$ & 34.6/27 \\
160 & 11470 & 1.82 & $6.5 \pm 2.8$ & $3.6 \pm 1.5$ & 20.2\textsuperscript{*} & \nodata & $-13.81 \pm 0.18$ & \nodata \\
185 & 11474 & 1.82 & $<6.4$ & $<3.5$ & 20.6\textsuperscript{*} & \nodata & $<-13.81$ & \nodata \\
211 & 11482 & 1.95 & $38.0 \pm 6.5$ & $5.8 \pm 1.9$ & $<21.4$ & $1.70 \pm 0.24$ & $-13.00 \pm 0.16$ & 29.4/30 \\
213 & 11483 & 1.94 & $34.0 \pm 6.2$ & $4.7 \pm 1.7$ & $<21.4$ & $2.20 \pm 0.30$ & $-13.25 \pm 0.18$ & 16.2/22 \\
214 & 11484 & 1.93 & $<5.9$ & $<3.0$ & 20.8\textsuperscript{*} & \nodata & $<-13.86$ & \nodata \\
215 & 11485 & 1.99 & $36.1 \pm 6.3$ & $3.4 \pm 1.4$ & $<21.1$ & $1.79 \pm 0.27$ & $-13.09 \pm 0.18$ & 28.7/27 \\
\hline
\multicolumn{9}{c}{Archival VLA sample}\\
\hline
47 & 11446 & 1.98 & $59.1 \pm 8.3$ & $17.6 \pm 3.3$ & 20.5\textsuperscript{*} & $0.36 \pm 0.12$ & $-12.27 \pm 0.09$ & 84.9/52 \\
69 & 13858 & 14.87 & $1444.1 \pm 41.0$ & $18.7 \pm 1.2$ & $<18.8$ & $1.97 \pm 0.11$\textsuperscript{\dag} & $-12.49 \pm 0.04$ & 145.0/104 \\
87 & 13859 & 14.88 & $266.9 \pm 17.5$ & $10.4 \pm 0.9$ & $22.0 \pm 0.1$ & $1.87 \pm 0.10$ & $-12.69 \pm 0.05$ & 84.0/105 \\
106 & 11456 & 1.82 & $208.2 \pm 15.4$ & $37.8 \pm 5.0$ & $<20.3$ & $1.56 \pm 0.11$ & $-12.19 \pm 0.06$ & 73.5/84 \\
146 & 5664 & 4.66 & $67.5 \pm 8.8$ & $6.9 \pm 1.3$ & $<20.2$ & $1.34 \pm 0.15$ & $-12.97 \pm 0.10$ & 94.4/53 \\
174 & 13863 & 14.92 & $791.3 \pm 30.3$ & $13.0 \pm 1.0$ & $<20.1$ & $1.92 \pm 0.06$ & $-12.67 \pm 0.03$ & 143.9/182 \\
203 & 11479 & 1.94 & $1610.3 \pm 43.1$ & $167.1 \pm 10.2$ & $<19.5$ & $2.25 \pm 0.05$ & $-11.65 \pm 0.03$ & 156.4/118\\
\enddata
\tablecomments{Column (1): \citetalias{GH07s} ID. Column (2): Chandra observation ID. Column (3): Chandra observation exposure time. Column (4): net counts for the full 0.5--7~keV band. Upper limits are given for non-detections at the 99\% confidence level \citep{Kraft91}. Column (5): full-band count rate. Column (6): logarithm of the best-fit column density (see Section \ref{subsec:Xobs}). Column (7): X-ray photon index, where $N_E \propto E^{-\Gamma}$. Error bars are given at the 1$\sigma$ level. Blank values indicate X-ray non-detections or too few counts for spectral fitting. Column (8): logarithm of unabsorbed hard-band (2--10~keV) flux. Error bars are given at the 1$\sigma$ level, while upper limits are given for non-detections at the 99\% level. Column (9): Cash statistic and degrees of freedom.\\
\textsuperscript{*} For objects with no spectral fit performed ($<10$ counts), we instead list $N_{\rm H,gal}$, the Galactic column density along the line of sight from the HI4PI survey \citep{HI4PI}.\\
\textsuperscript{\dag} X-ray spectral fit required the addition of a multi-temperature accretion disk component (see Section \ref{subsec:Xobs}).
}
\end{deluxetable*}

For X-ray sources with $>$10 full-band counts (12 objects, all of which are also detected in the hard 2--7~keV band), we extracted spectra using {\tt specextract} over 5$\arcsec$ apertures, and we performed spectral fitting using the Interactive Spectral Interpretation System (ISIS) v1.6.2 \citep{Houck00}. Since the majority of our spectra (9/12) have relatively few counts ($<$300), we binned each spectrum to 1 count per bin and performed our fitting using Cash statistics \citep{Cash79}. We fit each spectrum using an absorbed power-law model, {\tt tbabs*ztbabs*powerlaw}, where {\tt tbabs} was frozen to the Galactic column densities ($N_{\rm H,gal}$) from the HI4PI survey \citep{HI4PI}, and {\tt ztbabs} was left as a free parameter to estimate intrinsic absorption ($N_{\rm H,fit}$) at the redshift of each object. We did not attempt joint spectral fitting because parameters like $\Gamma$ could vary on a per-object basis. This provided acceptable fits for 11/12 spectra, and we did not pursue more complicated models for these 11 spectra. For the final spectrum, GH~69 (ObsID 13858; also see Appendix \ref{sec:app}), we found the addition of a multi-temperature accretion disk was sufficient, {\tt tbabs*ztbabs*(diskbb+powerlaw)}. From these 12 spectral fits, we calculated unabsorbed model fluxes over the 2--10~keV hard X-ray band using the {\tt cflux} convolution model in ISIS.

For objects with fewer than 10 full-band counts (including non-detections), we used the full-band count rates (or limits) to estimate unabsorbed hard-band fluxes (or upper limits) in the Portable, Interactive Multi-Mission Simulator (PIMMS).\footnote{We utilized the PIMMS v4.12a web tool hosted by the Chandra X-Ray Center at \href{https://cxc.harvard.edu/toolkit/pimms.jsp}{https://cxc.harvard.edu/toolkit/pimms.jsp}.} We assumed a power-law model with $\Gamma=1.9$ (from the weighted mean of the best-fit $\Gamma$ values from the 12 spectral fits), and we adopted $N_{\rm H,gal}$ from the HI4PI survey. 

Finally, we re-estimated the optical--X-ray properties of our sample using the following methods. We found $l_{\rm 2 keV}$ via hard-band (2--10~keV) fluxes or limits in order to characterize hard (coronal) X-ray emission and avoid possible contamination from a soft excess \citep[e.g.,][]{Done12, Dong12a, Ludlam15}. Where available, we used the best-fit values of $\Gamma$ in our calculations, otherwise we used the weighted sample mean ($\Gamma=1.9$). Following prior studies (e.g., see Section 3.2 of \citealt{Plotkin16}), we used \ha\ line luminosities ($L_{\rm H\alpha}$) from \citetalias{GH07s} along with the relationship $L_{\rm H\alpha} = 5.25 \times 10^{42}\, (L_{\rm 5100}/10^{44}\ {\rm erg~s}^{-1})^{1.157}$~erg~s$^{-1}$ from \citet{Greene05} to avoid possible host-galaxy contamination and find the AGN continuum luminosity at 5100~{\AA} ($L_{\rm 5100}$). We then assumed a power-law continuum following the form $f_{\nu} \propto \nu^{-0.44}$ \citep[][]{VB01} to estimate $l_{\rm 2500}$. We finally calculated $\aox$, $\daox$, and $\fweak$ following Eqs.\ (1--3), and as previously noted, we report these values in Table \ref{tab:sample} instead of the prior literature values.\footnote{Our re-reduced values are generally consistent with those from the literature, with an average change in $\aox$ of only 0.06. However, the detection status of three objects changed: two prior non-detections are now redetermined as faint detections and one prior detection is redetermined as a non-detection (see Appendix \ref{sec:app}).}

\subsection{New and Archival VLA Radio Observations} \label{subsec:radio_obs}

The 11 new radio observations were completed in X-band (utilizing the 3-bit samplers to cover 8--12~GHz) with the VLA in B configuration under Project ID 20A-292 (PI Plotkin) from 2020 June 23 to July 11. Observation details are tabulated in Table \ref{tab:obslog}. Each observation included a scan of a standard flux/bandpass calibrator, followed by scans of the science target interleaved between scans of a secondary phase calibrator.

We downloaded the uncalibrated Science Data Model-Binary Data Format (SDM-BDF) dataset for each target using the National Radio Astronomy Observatory (NRAO) Archive Access Tool and performed standard automated flagging, calibration, and weighting using the Common Astronomy Software Applications (CASA; \citealt{CASA}) package version 6.4.1 with VLA calibration pipeline version 2022.2.0.64. After reviewing the calibrated measurement set, we performed additional flagging as necessary. 

Data were imaged using the CASA task \texttt{tclean} with a typical resolution (synthesized beamwidth) $\approx$\,0$\farcs$50. We utilized the Multi-Scale, Multi-Term Multi-Frequency Synthesis deconvolution algorithm (\texttt{deconvolver\,=\,`mtmfs'}; \citealt{Rau11}) with \texttt{nterms\,=\,2} and, for objects that appeared to exhibit extended emission, \texttt{scales\,=\,[0,5,15]}. To suppress sidelobes, we used Briggs weighting \citep{Briggs95} with \texttt{robust} parameter values ranging from 0.5 to 2.0 depending on field crowding and sensitivity needs for fainter targets. We stopped the cleaning process at a threshold of 2$\times$ the background rms noise ($\sigma_{\rm rms}$) using \texttt{nsigma\,=\,2.0}.

The 7 archival radio observations were gathered from Project ID SD0129, completed 2012 December 13--16, and Project ID 12B-064, completed 2012 October 25 through 2013 January 6 (see \citealt{Gultekin14}). These observations were performed in X-band (utilizing the 8-bit samplers to cover 8--10~GHz) with the VLA in A configuration. For consistency with the new VLA sample, we downloaded uncalibrated SDM-BDF datasets and re-imaged the archival objects (including flagging, calibration, reduction, and cleaning) using the same CASA version and methods described above. We found a typical resolution $\approx$\,0$\farcs$20.

In each new and archival image, we estimated $\sigma_{\rm rms}$ from a source-free sky region in each image. $\sigma_{\rm rms}$ was found to be consistent with noise predicted by the VLA Exposure Calculator tool. We then used the CASA task \texttt{imfit} to measure flux densities at the center of each target galaxy by fitting at least one two-dimensional Gaussian. For target objects with $> 3\sigma_{\rm rms}$ detections, we extracted peak and integrated flux densities, central frequencies, detection coordinates, and beam-deconvolved component sizes (if extended) from \texttt{imfit}. We adopted the associated uncertainties as reported by \texttt{imfit}. For non-detections, we report upper flux density limits at the $3\sigma_{\rm rms}$ level. The radio properties of our sample are listed in Table \ref{tab:obslog}. We provide the full set of VLA images in Appendix \ref{sec:app}.

\begin{deluxetable*}{lccccccccc}[t]
\tablenum{3}
\tablecaption{VLA X-band Observations and Sample Radio Properties\label{tab:obslog}}
\tablehead{
\colhead{GH ID} & \colhead{TOS} & \colhead{Phase Cal} & \colhead{$\nu_{\rm cent}$} & \colhead{Robust} & \colhead{$\theta_{\rm maj}$} & \colhead{$r_{\rm maj}$} & \colhead{$S_{\rm p}$} & \colhead{$S_{\rm i}$} & \colhead{$\alphar$} \\
\colhead{} & \colhead{(s)} & \colhead{(SDSS J)} & \colhead{(GHz)} & \colhead{} & \colhead{(mas)} & \colhead{(pc)} & \colhead{(mJy bm$^{-1}$)} & \colhead{(mJy)} & \colhead{} \\
\colhead{(1)} & \colhead{(2)} & \colhead{(3)} & \colhead{(4)} & \colhead{(5)} & \colhead{(6)} & \colhead{(7)} & \colhead{(8)} & \colhead{(9)} & \colhead{(10)} 
}
\startdata
\multicolumn{10}{c}{New VLA sample (B configuration)}\\
\hline
25 & 2763 & 0339$-$0146\textsuperscript{a} & 9.8 & 2.0 & \nodata & \nodata & $< 0.011$ & $< 0.011$ & \nodata \\
73 & 891 & 0921$+$6215\textsuperscript{a} & 10.0 & 0.5 & point & $< 257$ & $0.196 \pm 0.007$ & $0.210 \pm 0.013$ & $0.6 \pm 0.4$ \\
80 & 2643 & 0921$+$6215\textsuperscript{a} & 10.0 & 2.0 & point & $< 831$ & $0.024 \pm 0.003$ & $0.026 \pm 0.007$ & $0.7 \pm 1.4$\textsuperscript{\dag} \\
104 & 2676 & 1035$+$5628\textsuperscript{a} & 10.0 & 0.5 & point & $< 739$ & $0.020 \pm 0.004$ & $0.030 \pm 0.010$ & $-1.4 \pm 2.2$\textsuperscript{\dag} \\
157 & 1017 & 1309$+$1154\textsuperscript{b} & 10.0 & 2.0 & $1280 \pm 580$ & $686 \pm 311$ & $0.025 \pm 0.006$ & $0.068 \pm 0.023$ & $-2.6 \pm 3.4$\textsuperscript{\dag} \\
160 & 2763 & 1309$+$1154\textsuperscript{b} & 10.0 & 0.5 & point & $< 390$ & $0.087 \pm 0.004$ & $0.079 \pm 0.007$ & $-0.6 \pm 0.5$\textsuperscript{\dag} \\
185 & 2496 & 1505$+$0326\textsuperscript{b} & 9.8 & 1.5 & $1870 \pm 900$ & $818 \pm 934$ & $0.020 \pm 0.005$ & $0.059 \pm 0.019$ & $-4.7 \pm 4.8$\textsuperscript{\dag} \\
211 & 981 & 1635$+$3808\textsuperscript{b} & 10.0 & 2.0 & \nodata & \nodata & $< 0.025$ & $< 0.025$ & \nodata \\
213 & 1062 & 1609$+$2641\textsuperscript{b} & 10.0 & 0.5 & point & $< 410$ & $0.233 \pm 0.009$ & $0.246 \pm 0.016$ & $-1.1 \pm 0.5$\textsuperscript{\dag} \\
214 & 948 & 1617$+$0246\textsuperscript{b} & 9.8 & 0.5 & $224 \pm 83$ & $151 \pm 56$ & $0.360 \pm 0.009$ & $0.378 \pm 0.016$ & $-1.0 \pm 0.3$ \\
215 & 1011 & 1653$+$3945\textsuperscript{b} & 10.0 & 0.5 & \nodata & \nodata & $< 0.029$ & $< 0.029$ & \nodata \\
\hline
\multicolumn{10}{c}{Archival VLA sample (A configuration)}\\
\hline
47 & 3578 & 0830$+$2410\textsuperscript{a} & 9.0 & 0.5 & $123 \pm 3$ & $32 \pm 1$ & $0.635 \pm 0.004$ & $0.761 \pm 0.007$ & $-0.5 \pm 0.1$ \\
69 & 2085 & 0914$+$0245\textsuperscript{a} & 9.0 & 1.0 & $351 \pm 157$ & $433 \pm 194$ & $0.018 \pm 0.004$ & $0.032 \pm 0.010$ & $-0.9 \pm 2.5$\textsuperscript{\dag} \\
87 & 2085 & 1008$+$0730\textsuperscript{b} & 9.0 & 0.5 & $194 \pm 20$ & $144 \pm 15$ & $0.278 \pm 0.006$ & $0.375 \pm 0.012$ & $-1.6 \pm 0.5$\textsuperscript{\dag} \\
106 & 3581 & 1110$+$6028\textsuperscript{b} & 9.0 & 0.5 & $154 \pm 4$ & $52 \pm 2$ & $0.726 \pm 0.004$ & $0.897 \pm 0.008$ & $-1.1 \pm 0.2$ \\
146 & 3581 & 1229$+$0203\textsuperscript{b} & 9.0 & 0.5 & $125 \pm 16$ & $95 \pm 12$ & $0.309 \pm 0.004$ & $0.357 \pm 0.008$ & $-0.8 \pm 0.2$ \\
174 & 2148 & 1419$+$5423\textsuperscript{b} & 9.0 & 0.5 & $159 \pm 109$ & $189 \pm 129$ & $0.034 \pm 0.005$ & $0.046 \pm 0.011$ & $-0.1 \pm 2.1$\textsuperscript{\dag} \\
203 & 3581 & 1602$+$3326\textsuperscript{b} & 9.0 & 0.5 & $145 \pm 10$ & $45 \pm 3$ & $0.274 \pm 0.004$ & $0.368 \pm 0.008$ & $-0.7 \pm 0.3$\\
\enddata
\tablecomments{Column (1): \citetalias{GH07s} ID. Column (2): total VLA exposure time on science target (does not exclude exposure time ``lost" to flagging during the data reduction process). Column (3): phase calibrator (flux density calibrator is noted via superscript). Column (4): observation central frequency. Column (5): \texttt{robust} parameter value used for Briggs weighting (see Section \ref{subsec:radio_obs}). Column (6): for resolved detections, the major axis size of the primary emission source (deconvolved from the synthesized beam) as reported by \texttt{imfit}. Unresolved detections are noted with ``point" (see Section \ref{sec:results}). Blank values indicate non-detections. Typical synthesized beam sizes are $\approx$\,0$\farcs$50 in B configuration and $\approx$\,0$\farcs$20 in A configuration (per-target beam sizes are given in Appendix \ref{sec:app}). Column (7): maximum observed radial extent of the primary nuclear emission component (see Section \ref{sec:results}). Column (8): peak flux density of the primary nuclear component (upper limits for non-detections are 3$\sigma_{\rm rms}$). Flux density measurements are taken at approximately 10~GHz for the ``new" sample and 9~GHz for the ``archival" sample. Column (9): integrated flux density of the primary nuclear component. Column (10): in-band, peak-flux radio spectral index.\\
\textsuperscript{a} Observation used flux density calibrator 3C 147.\\
\textsuperscript{b} Observation used flux density calibrator 3C 286.\\
\textsuperscript{\dag} The measured spectral index fails our adopted uncertainty threshold and is not used in subsequent analyses (see Section \ref{sec:results}).
}
\end{deluxetable*}

We also estimated in-band spectral indices for all detected objects. First, we divided the full frequency bandpass of each target evenly into four spectral window bins ($\approx$\,1~GHz per bin for the new targets and $\approx$\,0.5~GHz per bin for the archival targets). Following the procedures outlined above, we produced four new partial-bandwidth images and emission fits per target, and measured the flux density of the target in each image. We then weighted the central-frequency peak flux densities of these four new measurements by their $1\sigma$ errors and used them to fit (via a $\chi^{2}$ minimization routine) a model power-law continuum of the form $S_{\nu} \propto \nu^{\alphar}$, where $S_{\nu}$ is the flux density and $\alphar$ is the reported X-band spectral index. To estimate uncertainty in the spectral indices, we used a Monte Carlo algorithm to generate a set of 1,000 mock spectra. For each spectral window bin, we varied the flux density via random sampling of a normal distribution (where the mean and standard deviation were set respectively to the observed flux density and uncertainty), and the central frequency via random sampling of a uniform distribution (defined by the bin bandwidth). Finally, we assigned $\pm\,1\sigma$ errors to $\alphar$ by re-fitting a power-law model to each simulated spectrum and finding the standard deviation of the resulting spectral index distributions.

\subsection{A Sample of NLS1s for Visual Comparison} \label{subsec:comp_samp}
To provide a visual basis for comparison, we include in some figures a sample of radio-quiet, high-Eddington ($\ledd \gtrsim 0.3$) narrow-line Seyfert 1 (NLS1) galaxies from \citet{Yang20}, with a comparable redshift range ($z < 0.17$) but a slightly higher average mass ($\langle \log M_{\rm BH} \rangle \approx 6.8$). For this comparison sample, we adopt published measurements: 5~GHz radio luminosities from \citet{Yang20} along with 2--10~keV X-ray fluxes and host-subtracted AGN continuum fluxes at 5100~{\AA} as reported by \citet{Wang13} (see additional references therein). We calculate $\daox$ as described in Section \ref{subsec:Xobs}. This comparison sample is almost entirely X-ray normal and is shown by \citet{Yang20} to display $\lrlx \sim 10^{-5}$--$10^{-4}$.

NLS1s have been found to occupy extreme positions in the \citet{Boroson92} Eigenvector 1 and other related parameter spaces \citep[e.g.,][]{Sulentic00, Grupe04}. They are thought to possess low-mass SMBHs ($M_{\rm BH} \sim 10^{6}$--$10^{8}~M_\odot$) at high (near- or super-Eddington) accretion rates, and they often display extreme X-ray properties such as weakness, variability, and spectral complexity (see, e.g., \citealt{Gallo18} and references therein). The overlap of \citetalias{GH07s} AGNs with some NLS1 characteristics has been investigated to moderate extent (see, e.g., \citetalias{GH07s}; \citealt{GH07b, Desroches09, Dong12a}), but it is still unclear how distinct the two populations are. In summary, some features of classical NLS1s are shared with the \citetalias{GH07s} sample, such as relatively narrow broad-line emission, high Eddington ratio (although for \citetalias{GH07s} this is likely a selection effect), and predominant radio-quietness. However, \citetalias{GH07s} objects also tend to have different optical \ion{Fe}{2} and [\ion{O}{3}] strength distributions and flatter $\aox$ on average (although this may be similar specifically to super-Eddington NLS1s). While NLS1s are not key to this paper, we discuss them in the following sections when relevant.

\section{Results} \label{sec:results}

Most objects in our sample show unambiguous compact radio emission consistent with a point source (with a peak-to-integrated flux density ratio $S_{\rm p}/S_{\rm i} \gtrsim 0.8$) at the SDSS coordinates. Three objects are not detected. Two targets, GH~157 and GH~160, show small, secondary extended components consistent with a jet or jet remnant morphology. Two other targets, GH~185 and GH~203, show faint but extensive emission surrounding the entire primary component, possibly correlated with the optical galaxy morphology. We take our measurements from the primary ``nuclear" component to characterize radio emission from AGN activity (per-object details are given in Appendix \ref{sec:app}), and we use the peak flux density, $S_{\rm p}$, to derive related quantities (e.g., luminosities) unless otherwise noted.

For objects that are resolved by \texttt{imfit}, we derive a spatial estimate of the maximum radial extent of the nuclear emission component ($r_{\rm maj}$) using the major axis FWHM of the beam-deconvolved component angular size ($\theta_{\rm maj}$) as reported by \texttt{imfit}. For unresolved sources, we instead use the major axis FWHM of the synthesized beam to place an upper limit on the maximum radius of the emission component. Values of $r_{\rm maj}$ and $\theta_{\rm maj}$ are reported in Table \ref{tab:obslog}. In A configuration we estimate $r_{\rm maj}$ ranging from approximately 0.03 to 0.43~kpc (average 0.14~kpc), and in B configuration we estimate a range of approximately 0.15 to 0.83~kpc (average 0.54~kpc). We note that the X-ray-weak portion of our sample was covered only using B configuration, meaning it has been preferentially observed at lower spatial resolution. However, as we discuss in Section \ref{subsec:mechanism}, we do not expect resolution or emission scale to have a significant impact on our results.

The in-band radio spectral indices ($\alphar$, given in Table \ref{tab:obslog}) we measured for many of our faintest targets are highly uncertain, yet we do find a number of reasonable estimates that are consistent with a power-law spectrum, primarily among the archival VLA sample. The weighted mean of the full sample is $\langle \alphar \rangle = -0.7 \pm 0.4$ (the error is the weighted standard deviation). When extrapolating our observed X-band flux densities to lower frequencies (5 or 1.4~GHz) for subsequent analysis, we use a measured value of $\alphar$ only if its associated uncertainty is less than or equal to the weighted standard deviation of the sample, otherwise we assume $\alphar = -0.7$. We explored the use of 3~GHz VLA Sky Survey (VLASS; \citealt{Lacy20}) quick-look images\footnote{We retrieved pre-processed VLASS quick-look continuum images from the NRAO archive server and took measurements using the Cube Analysis and Rendering Tool for Astronomy (CARTA; \citealt{CARTA}). We corrected flux densities for a known $\sim$\,8$\%$ systematic underestimation \citep{Lacy19}.} to provide supplemental broadband spectral index estimates and found the following: With a typical background $\sigma_{\rm rms} \approx 0.12$--$0.17$~mJy, 6 of the 7 ``archival" targets and only 2 of the 11 ``new" targets in our sample were detected by VLASS at $> 3\sigma_{\rm rms}$. We estimated power-law spectral indices using the peak flux densities or upper limits at 3~GHz and 9 or 10~GHz (for objects with a VLASS upper limit, we considered the result a constraint). Given the lower sensitivity of the VLASS observations, these two-point broadband estimates had higher uncertainty levels ($\sigma_{\alphar} \approx 0.5$ on average for the detected objects) than the in-band estimates ($\sigma_{\alphar} \approx 0.4$) and did not provide better constraints. However, the two versions were in general agreement, validating the approach we describe above.

Finally, because it is highly relevant to the discussion in later sections, we stress that the VLA X-band observations show most of our targets to be confidently radio-quiet per the conventional radio--optical definition, with $\Ro=f\textsubscript{5 GHz}/f\textsubscript{4400 {\AA}} < 10$ (values given in Table \ref{tab:compare}).\footnote{We estimate $f\textsubscript{4400 {\AA}}$ from \ha\ line luminosities using the same method described for $l_{\rm 2500}$ in Section \ref{subsec:Xobs}.} Two objects push slightly into the ``radio-intermediate" regime, showing $10 \lesssim \Ro \lesssim 35$. We do not exclude the radio-intermediate objects from subsequent analysis unless otherwise noted.

\subsection{Comparison of Radio and X-ray properties} \label{subsec:properties}

\begin{deluxetable}{lccccc}[t!]
\tablenum{4}
\tablecaption{Multiwavelength Properties\label{tab:compare}}
\tablehead{
\colhead{GH ID} & \colhead{$\log L_{\rm R}$}  & \colhead{$\log L_{\rm X}$}  & \colhead{$\log \Rx$} & \colhead{$\Ro$} &  \colhead{$\daox$}  \\
\colhead{} & \colhead{(erg~s$^{-1}$)}  & \colhead{(erg~s$^{-1}$)}  & \colhead{} & \colhead{} &  \colhead{}  \\
\colhead{(1)} & \colhead{(2)} & \colhead{(3)} & \colhead{(4)} & \colhead{(5)} & \colhead{(6)}
}
\startdata           
\multicolumn{6}{c}{New VLA sample}\\           
\hline           
25 & $< 37.36$ & 41.30 & $< -3.94$ & $< 0.46$ & $-$0.53 \\
73 & 36.95 & 41.50 & $-$4.55 & 0.46 & $-$0.46 \\
80 & 37.41 & $< 40.97$ & $> -3.56$ & 0.50 & $< -$0.66 \\
104 & 37.33 & $< 41.22$ & $> -3.89$ & 1.74 & $< -$0.41 \\
157 & 37.17 & 42.32 & $-$5.15 & 0.51 & $-$0.16 \\
160 & 37.84 & 41.16 & $-$3.32 & 2.20 & $-$0.53 \\
185 & 36.87 & $< 40.84$ & $> -3.97$ & 0.19 & $< -$0.68 \\
211 & $< 36.73$ & 41.42 & $< -4.68$ & $< 0.34$ & $-$0.40 \\
213 & 37.91 & 41.38 & $-$3.46 & 4.23 & $-$0.36 \\
214 & 38.61 & $< 41.19$ & $> -2.58$ & 33.56 & $< -$0.42 \\
215 & $< 37.34$ & 41.86 & $< -4.52$ & $< 0.42$ & $-$0.34 \\
\hline           
\multicolumn{6}{c}{Archival VLA sample}\\           
\hline           
47 & 37.78 & 41.89 & $-$4.11 & 8.59 & $-$0.35 \\
69 & 37.82 & 43.17 & $-$5.35 & 0.66 & 0.13 \\
87 & 38.49 & 42.46 & $-$3.97 & 6.39 & $-$0.09 \\
106 & 38.26 & 42.21 & $-$3.95 & 18.10 & $-$0.06 \\
146 & 38.59 & 42.20 & $-$3.61 & 4.05 & $-$0.33 \\
174 & 38.07 & 42.95 & $-$4.88 & 1.84 & 0.09 \\
203 & 37.64 & 42.68 & $-$5.04 & 0.45 & $-$0.03 \\
\enddata           
\tablecomments{Column (1): \citetalias{GH07s} ID. Column (2): logarithm of radio luminosity $L_{\rm R} = \nu L_{\nu}$ at 5~GHz. Column (3): logarithm of 2--10~keV luminosity. Column (4): logarithm of the ratio $\Rx = \lrlx$. Column (5): radio-to-optical flux density ratio $\Ro=f\textsubscript{5 GHz}/f\textsubscript{4400 {\AA}}$, where $f\textsubscript{\rm 5 GHz}$ and $f\textsubscript{4400 {\AA}}$ are, respectively, the radio and optical flux densities at 5~GHz and 4400~{\AA} \citep[e.g.,][]{Kellermann89,Stocke92}. Column (6): X-ray deviation parameter (more negative values are X-ray weaker), repeated from Table \ref{tab:sample} for convenience.
}
\end{deluxetable}

\begin{figure*}[t!]
\gridline{\fig{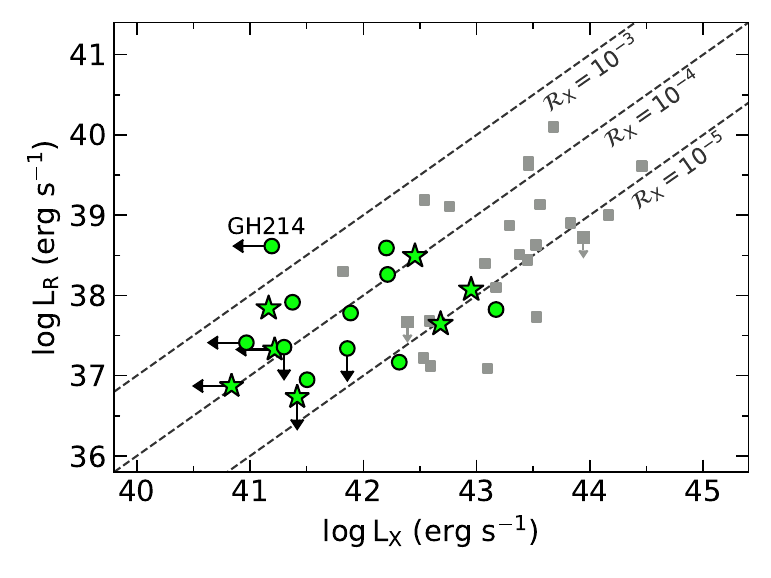}{0.49\textwidth}{\textbf{(a)}}
          \fig{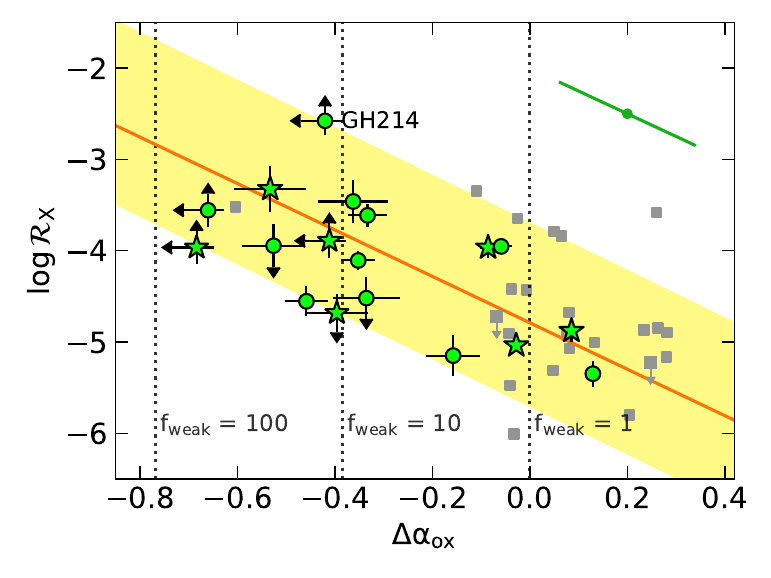}{0.49\textwidth}{\textbf{(b)}}}
\caption{\textbf{(a):} In log scale, 5~GHz radio luminosity, $L_{\rm R}$, vs.\ 2--10~keV luminosity, $L_{\rm X}$ (see Section \ref{subsec:properties}). Diagonal dashed lines show constant $\Rx = \lrlx$ at $10^{-5}$, $10^{-4}$, and $10^{-3}$. Green-filled symbols show our sample of \citetalias{GH07s} AGNs, and grey squares show the comparison sample from \citet{Yang20}. Detection limits are denoted with arrows. Throughout the remainder of this paper, circles show objects consistent with Seyfert-like Type 1 AGNs on the Baldwin-Phillips-Terlevich (BPT, \citealt{BPT}) diagram of their narrow emission lines, and stars show objects consistent with composite galaxies (see Section \ref{subsec:star_formation}).
\textbf{(b):} $\log \Rx$ vs.\ $\daox$ (see Section \ref{subsec:properties}). Vertical dashed lines mark values of $\fweak$ from 1 to 100 (leftward objects are X-ray weaker). Symbol definitions are the same as (a), except 1$\sigma$ error bars are included for our sample. The solid orange line shows the linear anticorrelation fit to our sample (note this fit \textit{does not} include the comparison sample). The green dot and diagonal bars in the upper right illustrate the potential influence of X-ray variability using the mock uncertainty scheme described in Section \ref{subsec:properties}. The yellow shaded area shows the model prediction described in Section \ref{subsec:weakness} (this is not the uncertainty for the orange linear fit). In both (a) and (b), the radio-intermediate object GH~214 is tagged by name (see Appendix \ref{sec:app}).
\label{fig:main_result}}
\end{figure*}

Our primary interest lies in evaluating the relationship between radio and X-ray emission. Following \citet{Terashima03}, we define the ratio of radio to X-ray luminosities as $\Rx = \lrlx$, where $L_{\rm R}$ is the radio luminosity $\nu L_{\nu}$ at $5$~GHz and $L_{\rm X}$ is the 2--10~keV X-ray luminosity. These quantities are listed in Table \ref{tab:compare} along with $\Ro$ and, for convenience, $\daox$. Figure \ref{fig:main_result}(a) shows where our sample resides in $L_{\rm R}$--$L_{\rm X}$ space. We find that the mean value of $\Rx$ in our sample of \citetalias{GH07s} AGNs ($\langle \log \Rx \rangle = -4.1 \pm 0.7$; the quoted error is the standard deviation) is higher than the comparison sample ($\langle \log \Rx \rangle = -4.7 \pm 0.8$) by $\sim$\,0.6~dex, although this does not appear particularly significant given the range of uncertainty.

Comparison with $\daox$ allows us to better evaluate $\Rx$ in terms of X-ray weakness, as shown in Figure \ref{fig:main_result}(b). Following the Monte Carlo approach described below (in order to account for error and non-detections),\footnote{The tests available in the Astronomical SURVival Statistics package (\texttt{ASURV}; \citealt{asurv2, asurv4}) assume that censored data follow intrinsic distributions consistent with the non-censored data. This may not be a realistic assumption in our case, so we adopt a Monte Carlo approach instead.} we observe a tentative anticorrelation between $\log \Rx$ and $\daox$ in our sample, finding a Kendall's Tau rank correlation coefficient $\tau = -0.48 \pm 0.05$ with a null-hypothesis probability $p_{\rm null} = 0.008 \pm 0.006$. A linear regression gives slope and intercept coefficients $m = -2.5 \pm 0.3$, $b = -4.8 \pm 0.1$ (where it is assumed $\log \Rx = m\daox + b$; solid orange line in Figure \ref{fig:main_result}(b)).

To find the above result, we use the following Monte Carlo scheme to generate $10^4$ iterations of mock $\log \Rx$ and $\daox$ measurements for each object. For radio, we randomly vary the X-band radio flux density and spectral index $\alphar$ via normal distributions (the mean and standard deviation for the former are the observed flux density and uncertainty, while for the latter we use either the measured slope and error or sample weighted mean and error according to the rule defined in Section \ref{sec:results}) before recalculating $L_{\rm R}$. For the X-ray, we similarly vary the flux and photon index $\Gamma$ via normal distributions (the mean and standard deviation are again the observed flux or photon index and the associated uncertainty) before recalculating $L_{\rm X}$ and $\daox$. For radio or X-ray non-detections, flux values are drawn randomly from a uniform distribution between the reported upper limit and 50\% of the upper limit, and we incorporate uncertainty on our assumed $\alphar$ or $\Gamma$ values by using the sample weighted mean and standard deviation to define the normal distribution for their random variation. For each Monte Carlo iteration, we perform a linear fit via orthogonal distance regression (weighted by uncertainty) and Kendall's Tau correlation test; the results reported above give the mean and $\pm\,1\sigma$ error (found from the 16\textsuperscript{th} and 84\textsuperscript{th} percentile values) of the distributions from these tests. We also assign $\pm\,1\sigma$ error bars to each data point in Figure \ref{fig:main_result}(b) by finding the 16\textsuperscript{th} and 84\textsuperscript{th} percentile values of each object's mock $\Rx$ and $\daox$ distributions.

To better determine the impact of certain choices we have made in defining $\Rx$, we produce variants of the $\log \Rx$--$\daox$ fit as follows. While these alternative results are largely consistent with the above error analysis, we describe them briefly here and report them along with the primary result in Table \ref{tab:fits} for completeness. In the primary fit reported above (v.1 in Table \ref{tab:fits}), $L_{\rm R}$ is found using estimated 5~GHz flux densities; to eliminate the error possible from extrapolating along an uncertain spectral index, we produce an alternative and slightly more robust fit (v.2) with $L_{\rm R}$ taken at the observed central frequency ($\nu_{\rm cent} \sim$ 9--10~GHz depending on target; see Table \ref{tab:obslog}). We also produce a version of this fit (v.3) using integrated core-component flux densities to show that our results are not significantly influenced by our primary choice of peak flux densities, as well as a version (v.4) that excludes all 7 objects with radio and/or X-ray non-detections (although for this last we point out the small remaining sample size and high uncertainty).

\begin{deluxetable}{lccc}[t!]
\tablenum{5}
\tablecaption{$\log \Rx$--$\daox$ Linear Fit \label{tab:fits}}
\tablehead{
\colhead{Fit Version} & \colhead{$m$}  & \colhead{$b$}  
& \colhead{$p_{\rm null}$} \\
\colhead{(1)} & \colhead{(2)} & \colhead{(3)} & \colhead{(4)} 
}
\startdata
v.1: $L_{\rm R}$ ($5$~GHz) & $-2.5 \pm 0.3$ & $-4.8 \pm 0.1$ 
& $0.008 \pm 0.006$ \\
v.2: $L_{\rm R}$ ($S_{\rm p}$ at $\nu_{\rm cent}$) & $-2.8 \pm 0.2$ & $-4.8 \pm 0.1$ 
& $0.006 \pm 0.004$ \\
v.3: $L_{\rm R}$ ($S_{\rm i}$ at $\nu_{\rm cent}$) & $-2.8 \pm 0.2$ & $-4.7 \pm 0.1$ 
& $0.004 \pm 0.003$ \\
v.4: Excl. non-det.\textsuperscript{*} & $-3.3 \pm 0.5$ & $-4.8 \pm 0.1$ 
& $0.045 \pm 0.036$ \\
v.5: Excl. $\Ro$-int.\textsuperscript{*} & $-2.4 \pm 0.2$ & $-4.9 \pm 0.1$ 
& $0.009 \pm 0.007$ \\
\enddata
\tablecomments{Column (1): version of $\log \Rx$--$\daox$ linear fit (for v.1--4 see Section \ref{subsec:properties}; for v.5 see Section \ref{subsec:mechanism}). Column (2): fit slope. Column (3): fit y-intercept. 
Column (4): null-hypothesis probability from Kendall's Tau rank correlation test.\\
\textsuperscript{*} $L_{\rm R}$ at 5~GHz.
}
\end{deluxetable}

While the above simulations suggest that none of the data points in Figure \ref{fig:main_result}(b) has substantial error (apart from non-detections), our ability to fully account for uncertainty is likely complicated by other phenomena or relationships (e.g., co-dependence of $\Gamma$ and $\daox$). The influence of X-ray variability is particularly challenging to assess, as variability estimates are only available for a few of our targets (discussed in Section \ref{subsec:variability}). However, we do attempt a rudimentary illustration of the effect of X-ray variability. We follow the Monte Carlo scheme described above for X-ray uncertainty, but incorporate an additional random variation via normal distribution where both the mean and the standard deviation are defined by the mock flux value. The green point and diagonal bars in the upper right of Figure \ref{fig:main_result}(b) illustrate the average effect X-ray variability might have on individual uncertainties.

Finally, we comment briefly on whether there may be a redshift or evolutionary dependence in our results. \citet{Rankine23} show that the X-ray luminosity distribution across redshift may have an impact on $\aox$ (at least in SDSS quasars, with $M_{\rm BH}$ potentially being the driver of the evolution). While this may be the reason for finding generally flatter values of $\aox$ in mBHs \citep[e.g.,][]{Desroches09, Baldassare17}, we stress that our sample covers a fairly narrow range in both mass and redshift. We find no correlation or dependency between combinations of redshift or mass and $\Rx$, $\Ro$, or $\daox$ in our sample ($p_{\rm null} \sim 0.6$).

\section{Discussion} \label{sec:discussion}

Prior empirical correlations indicate that $\Rx \sim$ constant (albeit with often substantial scatter of up to $\sim$\,2~dex) in $\Ro$-quiet,\footnote{Hereafter, in order to easily differentiate between optical and X-ray modes of radio loudness, we utilize the $\Ro$ and $\Rx$ nomenclature. I.e., we refer to radio-quiet vs.\ optical as ``$\Ro$-quiet," radio-intermediate vs.\ X-ray as ``$\Rx$-intermediate," and so forth.} X-ray-normal AGNs (e.g., \citealt{Panessa07, Panessa15, Panessa22, Laor08, Gultekin14, Gultekin22, Behar15, Smith20, Yang20, Chen23}). As described in the previous section, we observe a tentative but intriguing anticorrelation between $\Rx$ and $\daox$ in our sample, implying that as X-ray emission is weakened, $\Rx$ loudness is increasing rather than remaining constant. We now discuss possible physical interpretations for this observation.

\subsection{Impact of the AGN Radio Emission Mechanism(s)} \label{subsec:mechanism}

Before discussing possible sources of X-ray weakness, we investigate how our results may be affected by the source of radio emission. While $\Ro$-loud AGNs are typically seen to be dominated by jets and diffuse radio lobes, $\Ro$-quiet AGNs often show signatures of a number of alternative radio components at various physical scales (e.g., \citealt{Panessa19} and references therein; \citealt{McCaffrey22, Yang22, Yang24}). Diagnosis of the dominant emission mechanism can sometimes be informed by the radio spectral index $\alphar$. At the physical scale and frequency pertinent to our observations, a compact corona or jet base is expected to have $\alphar \gtrsim -0.5$ from optically thick, self-absorbed synchrotron emission; non-collimated outflows such as disk winds may produce $\alphar \approx -0.1$ from optically thin free-free emission; shocks related to jet lobes or disk winds will show $\alphar \lesssim -0.5$ from optically thin synchrotron emission; and star formation will also often result in $\alphar \lesssim -0.5$. For our purposes, we define X-band spectral slopes $\alphar > 0$ as ``inverted," $-0.5 \leq \alphar \leq 0$ as ``flat," and $\alphar < -0.5$ as ``steep."

For $\Ro$-quiet, supermassive AGNs, the approximation that $\Rx \sim$\,constant seems to hold even at kpc scales (e.g., from their results at 1.4~GHz, \citealt{Panessa15} suggest that even the total, large-scale radio emission is linked to accretion-related X-ray emission), yet the exact value of the relationship and the extent of its scatter will be at least partly dependent on the dominant radio emission mechanism being observed \citep[see, e.g.,][]{Laor08, Behar15, Yang20, Panessa22}. We also reiterate that the exact nature of the relationship is not yet understood. \citet{Chen23} examined Very Long Baseline Array observations of a sample of $\Ro$-quiet quasars in which radio emission at physical scales $\lesssim 0.4$~pc displays generally flat/inverted spectral slopes and follows $\log \Rx \sim -6$ with a tight correlation, and they noted that VLA observations of the same sample (at larger physical scales) show generally steeper slopes and follow $\log \Rx \sim -5$ with more scatter in the correlation. As they explain, this is consistent with small-scale (sub-pc) nuclear emission being produced by a compact, optically thick core component (e.g., corona or jet base), and larger-scale (sub-kpc) nuclear emission being associated with a variety of possible extended components. Similar results recently obtained by \citet{Wang23} and \citet{Shuvo23} appear to corroborate this interpretation, suggesting that an anticorrelation or bimodal relationship between $\alphar$ and $\Rx$ might be observable for X-ray-normal objects. Additionally, \citet{Laor19} and \citet{Alhosani22} show that radio slope and compactness are anticorrelated with Eddington ratio (i.e., they find compact emission with flat/inverted slopes at $\ledd < 0.3$, and less-compact emission with steep slopes at $\ledd> 0.3$) in $\Ro$-quiet AGNs. These observations are consistent with a unified accretion model in which nuclear outflows are more likely at high $\ledd$ \citep[e.g.,][]{Giustini19}. 

\begin{figure*}[t]
\gridline{\fig{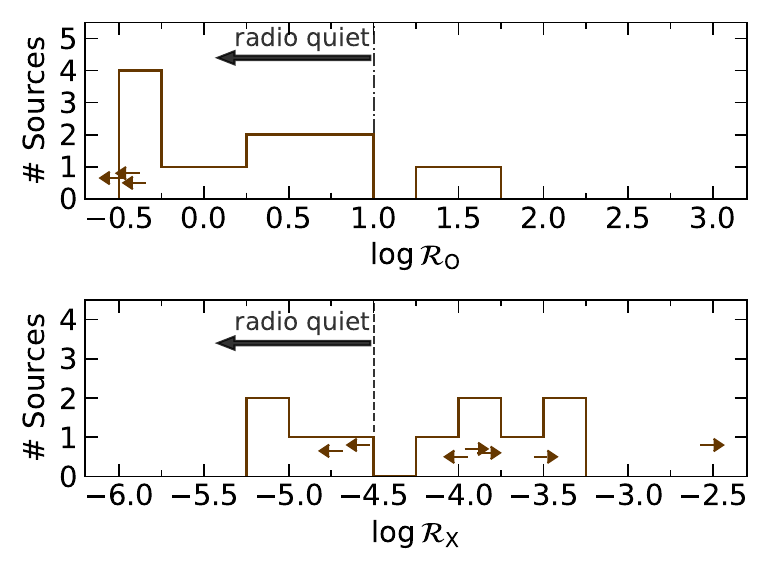}{0.49\textwidth}{\textbf{(a)}}
        \fig{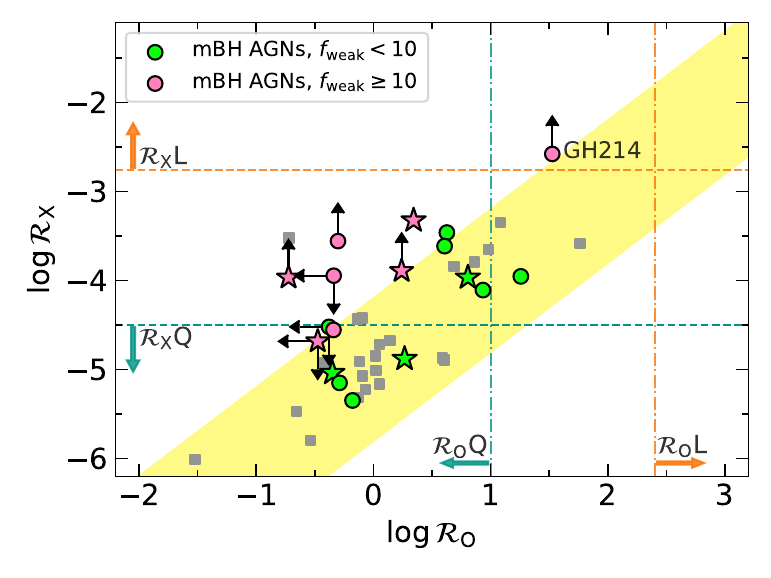}{0.49\textwidth}{\textbf{(b)}}}
\caption{\textbf{(a):} Histograms comparing our sample's distributions of radio loudness parameters, $\Ro$ (top) and $\Rx$ (bottom), on matched log scales. Arrows denote limits from non-detections. The dash-dot and dashed vertical lines respectively show our adopted $\Ro$-quiet and $\Rx$-quiet demarcations. Objects to the left of these lines are considered radio-quiet. Our sample is predominantly radio-quiet in $\Ro$ but not $\Rx$, suggesting that X-ray is preferentially weakened.
\textbf{(b):} Comparison of $\Rx$ and $\Ro$ in log scale. Symbols are as defined for Figure \ref{fig:main_result}(a), except objects in our sample with $\fweak \geq 10$ are pink-filled and those with $\fweak < 10$ are green-filled. The vertical dash-dot lines show the conventional $\Ro$-quiet (teal) and $\Ro$-loud (orange) demarcations, and the horizontal dashed lines show our adopted $\Rx$-quiet (teal) and $\Rx$-loud (orange) demarcations (based on empirical boundaries suggested by \citealt{Terashima03} and \citealt{Panessa07}). The yellow-shaded area roughly illustrates the tendency for large, radio-diverse samples of AGNs to experience a tandem increase in radio loudness on both axes (see, e.g., Figure 5 of \citealt{Zuther12}). While our sample is largely $\Ro$-quiet, the $\fweak \geq 10$ objects deviate from the trend further into $\Rx$-intermediate space, suggesting that X-ray is preferentially weakened. \citet{Zuther12} note that this behavior is also seen with broad absorption line quasars (which may experience X-ray weakness due to obscuration). 
\label{fig:Radio_loud}}
\end{figure*}

From all of the above, we can infer that the scatter of $\Rx$ in our sample may be partly due to varied nuclear radio emission mechanisms. Indeed, as with many of the above-noted studies, our sample spans a range of roughly 2~dex in $\Rx$ (see Figure \ref{fig:main_result}(a)). While it may be possible for X-ray-normal populations to show an $\alphar$--$\Rx$ relationship as described above, in an X-ray shielding scenario it is likely that the X-ray-weak objects would deviate from such a relationship, with some compact, flat/inverted-slope radio sources moving into the higher $\Rx$ range. Unfortunately, we lack a sufficient number of observations with the sensitivity or resolution needed to make statistical statements regarding the radio emission mechanisms of our sample. As our sample spans a fairly narrow range of high $\ledd$, it is also probably not meaningful to attempt to use $\ledd$ as a surrogate for $\alphar$ (and for completeness we note that a correlation between $\Rx$ and $\ledd$ is not found; $p_{\rm null} = 0.5 \pm 0.2$). More sensitive and higher resolution follow-up observations of the nuclear radio emission (from, for example, a next-generation VLA; e.g., \citealt{Selina18, Plotkin18}) will allow us to better understand the dominant radio emission mechanism and its relationship with X-ray emission \citep[e.g.,][]{Behar15, Behar20, Chen22}.

However, the radio emission mechanism is unlikely to be directly responsible for the observed $\Rx$--$\daox$ anticorrelation because, as we discuss below, it does not appear that $\Rx$ loudness in our sample can be explained simply by excess or boosted radio emission. First, we note that the $\log \Rx$--$\daox$ fit produced using integrated radio flux densities (v.3 in Table \ref{tab:fits}) is consistent with the versions derived from peak flux densities, which implies that while a generally larger observed radio scale for the X-ray-weak objects (from lower resolving power in VLA B configuration) may introduce a risk of observational bias in radio emission mechanisms, such a bias is unlikely to be responsible for the anticorrelation.

Next, the histograms shown in Figure \ref{fig:Radio_loud}(a) compare measures of radio loudness via the distributions of $\Rx$ and $\Ro$, and they suggest that a majority of our sample are comparatively weak in X-rays. Expanding on this observation, Figure \ref{fig:Radio_loud}(b) explores our sample on the $\log \Ro$--$\log \Rx$ plane, 
again with the $\Ro$-quiet NLS1s from \citet{Yang20} for visual comparison. The vertical dash-dot lines denote the conventional $\Ro$-quiet (teal; $\log \Ro=1$) and $\Ro$-loud (orange; $\log \Ro=2.48$) demarcations, and we refer to the range between as $\Ro$-intermediate. The horizontal dashed lines similarly show our adopted $\Rx$-quiet (teal; $\log \Rx = -4.5$) and $\Rx$-loud (orange; $\log \Rx = -2.75$) demarcations (based on empirical boundaries suggested by \citealt{Terashima03} and \citealt{Panessa07}), and we refer to the range between as $\Rx$-intermediate. Much of our sample resides in a clearly $\Rx$-intermediate space, and, importantly, the X-ray weakest ($\fweak \gtrsim 10$, shown by pink symbols) appear to have $\Rx$ values higher than typical compared to their $\Ro$ (see Figure \ref{fig:Radio_loud}(b) caption text pertaining to the yellow shaded band). We also find that excluding the two $\Ro$-intermediate objects from our $\Rx$--$\daox$ analysis still gives a consistent fit result (v.5 in Table \ref{tab:fits}).

Furthermore, we do not find evidence of a relationship between $\daox$ and any other parameter that, in the context of the above discussion, could be linked to the radio emission mechanism, including $r_{\rm maj}$, $\Ro$, $\ledd$, or $\alphar$ (acknowledging limited dynamic range in several parameters, and also very small number statistics for the $\alphar$ subsample).

\subsubsection{Relativistic Beaming} \label{subsec:beaming}

We find good indications that our radio measurements are not influenced by relativistic beaming from jets oriented along our line of sight. If they were, we might generally expect to find $\Ro$-louder objects with flat/inverted radio spectra ($\alphar \geq 0$) and boosted optical continua \citep[e.g.,][]{Blandford78, Urry95}. Most of our targets are decidedly $\Ro$ quiet, with, on average, a steep radio spectrum. A boosted optical spectrum has largely been ruled out by the presence of broad optical emission lines during AGN classification (\citetalias{GH07s}). The subset of targets that are $\Ro$ intermediate have reasonably robust in-band spectral indices showing steep radio spectra. Finally, while we do not illustrate it here, comparison of our sample to standard blazar diagnostic plots indicates that our sample is consistent with $\Ro$-quiet quasars \citep[e.g.,][]{Shemmer09, Plotkin10}.

\subsubsection{Star Formation} \label{subsec:star_formation}

\begin{figure}[t]
\epsscale{1.15}
\plotone{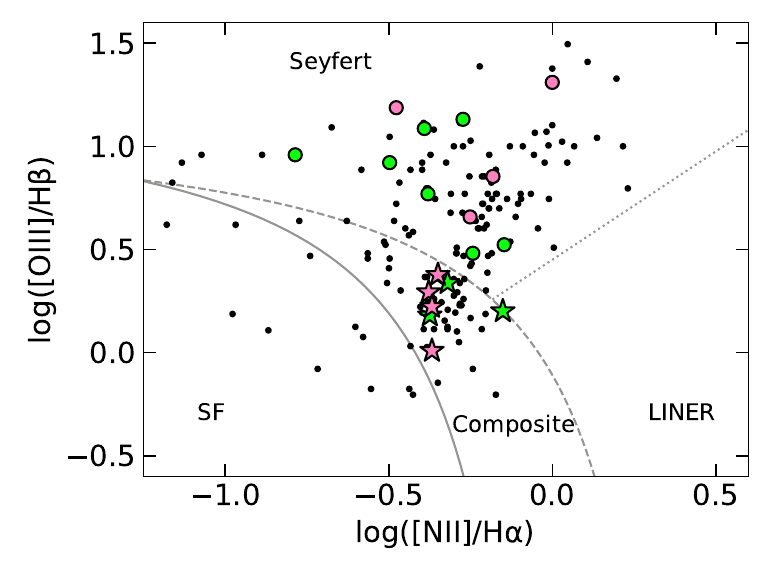}
\caption{BPT diagram \citep{BPT} comparing narrow emission line ratios (see Section \ref{subsec:star_formation}). Objects in our sample with $\fweak \gtrsim 10$ are pink-filled, and those with $\fweak < 10$ are green-filled. The black dots show the remainder of the \citetalias{GH07s} high-confidence parent sample (excluding objects with limits on one or more of the emission line strengths). The solid curved line shows the empirical ``pure star formation" demarcation from \citet{Kauffmann03}, while the curved dashed line shows the theoretical ``maximum starburst line" from \citet{Kewley01}; ``composite" galaxies fall between the two. The diagonal dotted line divides Seyferts and low-ionization nuclear emission-line regions (LINERs) \citep{Schawinski07}.
\label{fig:BPT_diag}}
\end{figure}

The majority of our sample shows optical narrow emission line ratios (values from \citetalias{GH07s}) consistent with Seyfert-like Type 1 AGN activity on the Baldwin-Phillips-Terlevich diagram (BPT, \citealt{BPT}; see also, e.g., \citealt{Kewley01, Kauffmann03, Schawinski07}). Seven objects are consistent with ``composite" galaxies, suggestive of a combination of AGN and star formation activity. The BPT classifications of our targets are shown in Figure \ref{fig:BPT_diag} and listed in Table \ref{tab:sf}. As our earlier figures illustrate, however, the composite galaxies (represented by star symbols) show a largely unbiased distribution. This provides a first indication that star formation does not directly influence our results. 

In radio, we expect AGN emission to dominate over star formation activity within the $\lesssim$\,1~kpc projected physical radius we have observed for each object, given that the relatively large SDSS aperture (3{\arcsec} fiber diameter) probes our sample at a wider projected physical radius ($\lesssim$\,5~kpc). As a test, we estimate the star formation rate needed to produce the observed radio emission. \citet{Ho05} examined the ratio of [\ion{O}{2}] $\lambda$\,3727 to [\ion{O}{3}] $\lambda$\,5007 emission line strengths in AGNs and suggested that a ratio [\ion{O}{2}]/[\ion{O}{3}] $\approx$ 0.1--0.3 is consistent with the observed [\ion{O}{2}] emission originating purely from the AGN's narrow-line region, while amounts in excess of that can be attributed to star formation. In our sample, we find [\ion{O}{2}]/[\ion{O}{3}] ratios spanning 0.11--1.2 (measurements obtained from \citetalias{GH07s}), with a mean $\approx$\,0.45, suggesting the presence of some excess [\ion{O}{2}] emission related to star formation.\footnote{We caution that \citetalias{GH07s} express concerns regarding use of [\ion{O}{2}] to estimate star formation activity. They note that while the colors and spectral properties of their sample suggest the presence of intermediate-age stellar populations, their data cannot directly constrain emission contributions from the AGN narrow-line regions. The \citet{Ho05} assessment may therefore not be appropriate for more rigorous evaluation of star formation activity in this sample.} 

For each object, we estimate an upper limit on the star formation rate from the [\ion{O}{2}] line luminosity using the updated calibrations of \citet{Zhuang19} (see Eq.~7 and related discussion/references therein). We then translate that estimate to an expected 1.4~GHz monochromatic luminosity by inverting Eq.~(12) of \citet{Kennicutt12} and applying the star formation rate calibration of \citet{Murphy11}. We also calculate the ``observed" 1.4~GHz monochromatic luminosity by converting the X-band integrated flux density ($S_{\rm i}$) using the measured or assumed spectral index ($\alphar$) per the rule set in Section \ref{sec:results}. Both the observed monochromatic luminosity, $L_{\rm 1.4}$, and the translated monochromatic luminosity expected from the [\ion{O}{2}]-derived star formation rate, $L_{\rm 1.4}$([\ion{O}{2}]), are presented in Table \ref{tab:sf}. We again stress that the SDSS aperture probes a larger region than the radio observations, meaning $L_{\rm 1.4}$([\ion{O}{2}]) can be treated as an upper limit. In almost all cases (including the BPT-composites), this [\ion{O}{2}]-derived upper limit is at least one order of magnitude lower than the observed radio luminosity. We infer from this analysis that while we cannot discount the possibility of a contribution from star formation, the compact radio emission we observe cannot be explained by star formation alone and is likely to be dominated by AGN activity.  

\begin{deluxetable}{lccc}[t]
\tablenum{6}
\tablecaption{Star Formation Analysis\label{tab:sf}}
\tablehead{
\colhead{GH ID} & \colhead{BPT}  & \colhead{$\log L_{\rm 1.4}$}  & \colhead{$\log L_{\rm 1.4}$([\ion{O}{2}])} \\
\colhead{} & \colhead{}  & \colhead{(erg~s$^{-1}$~Hz$^{-1}$)}  & \colhead{(erg~s$^{-1}$~Hz$^{-1}$)}  \\
\colhead{(1)} & \colhead{(2)} & \colhead{(3)} & \colhead{(4)} 
}
\startdata
\multicolumn{4}{c}{New VLA sample}\\
\hline
25 & Seyfert & $< 28.01$ & 27.00 \\
73 & Seyfert & 26.96 & 24.72 \\
80 & Seyfert & 28.11 & 26.65 \\
104 & composite & 28.16 & 27.58 \\
157 & Seyfert & 28.27 & 27.58 \\
160 & composite & 28.47 & 27.89 \\
185 & composite & 28.01 & 27.63 \\
211 & composite & $< 27.41$ & 26.79 \\
213 & Seyfert & 28.61 & 27.11 \\
214 & Seyfert & 29.44 & 25.95 \\
215 & Seyfert & $< 28.01$ & 26.70 \\
\hline
\multicolumn{4}{c}{Archival VLA sample}\\
\hline
47 & Seyfert & 28.41 & 27.09 \\
69 & Seyfert & 28.73 & 27.59 \\
87 & composite & 29.29 & 28.15 \\
106 & Seyfert & 29.25 & 28.16 \\
146 & Seyfert & 29.37 & 27.90 \\
174 & composite & 28.85 & 28.31 \\
203 & composite & 28.43 & 27.40
\enddata
\tablecomments{Column (1): \citetalias{GH07s} ID. Column (2): classification on the BPT diagram. Column (3): logarithm of the observed monochromatic radio luminosity at 1.4~GHz. Column (4): logarithm of the expected 1.4~GHz monochromatic luminosity as estimated from the [\ion{O}{2}]-derived star formation rate (see Section \ref{subsec:star_formation}).
}
\end{deluxetable}

Finally, we note that the multiwavelength properties (including the BPT classifications) and prior analyses of our sample are largely inconsistent with significant contributions from supernovae and/or their remnants. For example, supernova remnants are expected to be more $\Rx$-loud than observed in our sample (e.g., \citealt{Reines11} and references therein; however, see also \citealt{Hebbar19}). Furthermore, while supernova remnants could contribute broad H$\alpha$ emission or otherwise mimic AGN spectra \citep[e.g.,][]{Filippenko89, Reines13} in a way that increases perceived X-ray weakness,\footnote{As described in Section \ref{subsec:Xobs}, we use the broad H$\alpha$ luminosity in deriving $\daox$.} our X-ray-weak targets simply do not display the systematic strengthening of broad H$\alpha$ that would be needed to explain the observed $\Rx$--$\daox$ anticorrelation.

\subsection{Sources of X-ray Weakness} \label{subsec:weakness}

From the above analysis, we conclude that the increase in $\Rx$ loudness observed in our sample does not appear to be an artifact of enhanced radio loudness, and is instead related to genuine X-ray weakness. We defer discussion of X-ray variability to Section \ref{subsec:variability} and assume for the moment that our sample is not particularly variable. In the case of intrinsic X-ray weakness, the previously described empirical correlations suggest that $L_{\rm R}$ and $L_{\rm X}$ should scale in tandem, and that we should still generally expect the ratio $\Rx$ to remain constant, independent of $\daox$ (although we note that this expectation remains to be confirmed, with the challenge being identification of a purely intrinsically X-ray-weak sample with radio coverage).

On the other hand, if there is a compact shielding medium that obscures X-rays along our line of sight (while other emission regions remain unobscured) then we should see $\Rx$ anticorrelate with $\daox$. To examine what might be observed in this case, we construct a simple prediction model,
\begin{equation} \label{eq4}
\Rx = \fweak \Rxexp,
\end{equation}
where $\fweak$ is derived from $\daox$ as described in Section \ref{subsec:Xobs}, and $\Rxexp$ is the value of $\Rx$ that would otherwise be expected for the object from an X-ray-unobscured line of sight. Converting to $\log \Rx$--$\daox$ space,  
\begin{equation} \label{eq5}
\log \Rx = -2.6\daox + \log \Rxexp.
\end{equation}
We see that the predicted slope is $-2.6$, which is consistent with the best-fit slope from the linear regression to our sample ($-2.5 \pm 0.3$; see Section \ref{subsec:properties}). Our sample's best-fit intercept value is $-4.8\pm0.1$. However, the expected (unshielded) value of $\Rxexp$ for each AGN is likely to be partly dependent on the dominant radio emission mechanism (as discussed in Section \ref{subsec:mechanism}), so we adopt $\log \Rxexp = -4.7$ from the \citet{Yang20} sample mean and include a $\pm$~1 dex scatter. The yellow-shaded area in Figure \ref{fig:main_result}(b) illustrates this simple model, and while it is not conclusive, it appears at least qualitatively consistent with our observations. More sensitive observations of our X-ray-weakest targets and larger samples with X-ray-weak, high-Eddington AGNs will help overcome small-sample statistical limitations, improve dynamic range, and better test the prediction.

While it is possible to simply interpret the observed anticorrelation as being induced by the fact that $\daox$ and $\Rx$ are dependent on the X-ray luminosity and its inverse (respectively), we stress again that we take the $\sim$\,constant $\lrlx$ empirical relationship to suggest a null hypothesis of \textit{no} correlation expected between $\daox$ and $\Rx$. That is, for whatever reason, the mBH AGNs that appear X-ray weak relative to the optical do not follow the expected $\lrlx$ behavior derived from $\Ro$-quiet, X-ray-normal AGNs with X-ray luminosities spanning many orders of magnitude.

Indeed, other works have previously discussed the possibility of X-ray shielding/absorption in the context of $\Rx$ analysis involving higher-mass AGNs. For example, in their examination of the drivers of radio spectral slopes in  quasars, \citet{Laor19} note that a few of the objects in their sample have higher (X-ray-weaker) values of $\Rx$ and also show significant signs of X-ray absorption, including one confirmed broad absorption line (BAL) quasar.\footnote{By convention, narrow absorption lines (NALs) typically have FWHM $<500$ km s$^{-1}$, broad absorption lines (BALs) have FWHM $>2000$ km s$^{-1}$, and mini-BALs occupy the range between (see, e.g., \citealt{Weymann81}; \citealt{Hamann04} and references therein; \citealt{Gibson09}).} BAL quasars are often seen to be X-ray weak, an observation historically attributed to orientation-based obscuration \citep[e.g.,][]{Gallagher01, Gallagher02, Gallagher06}, and recent studies have further linked BAL activity and X-ray weakness/variability in high-Eddington BAL quasars to strong outflows \citep[e.g.,][]{Yi19a, Yi19b, Yi20, Rankine20, Saez21}. Likewise, \citet{Zuther12} find that BAL quasars which are radio-quiet per $\Ro$ still have high observed $\Rx$, in a manner similar to the X-ray-weak portion of our sample (see their Figure 5 in comparison to our Figure \ref{fig:Radio_loud}(b)). Still, we cannot ignore the possibility of intrinsic X-ray weakness in some objects \citep[e.g.,][]{Liu18BAL, Vito18}.

NLS1s are often found to possess complex X-ray spectra along with rapid and large-amplitude variability, which may suggest dynamic, possibly multi-component coronae that fluctuate with changes in accretion flow (e.g., \citealt{Gallo18} and references therein). Such properties imply that NLS1s which are X-ray weak are often intrinsically so. However, samples of NLS1s have been found with (albeit inconclusive) suggestions of X-ray absorption. \citet{Williams04} examined a sample of NLS1s in which the X-ray weakest objects tend to display flatter X-ray spectra with no sign of absorption in their optical spectra (which may suggest X-rays are preferentially obscured). Likewise, \citet{Gallo06} divided a sample of NLS1s into objects with ``simple" (i.e., power-law-consistent) and ``complex" X-ray spectra; they found that the complex-spectrum objects also tend to be significantly X-ray weaker with signs of absorption or reflection (as well as stronger optical \ion{Fe}{2} emission, which may indicate higher accretion rates in the context of Eigenvector 1).

\begin{figure}[t]
\epsscale{1.15}
\plotone{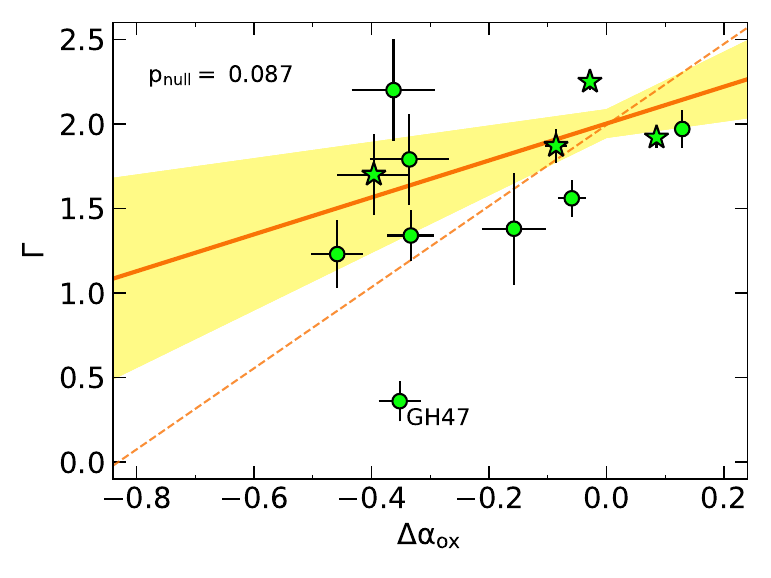}
\caption{X-ray photon index, $\Gamma$, vs.\ $\daox$ (see Section \ref{subsec:weakness}). Symbols are as defined for Figure \ref{fig:main_result}(a). Note that objects with X-ray non-detections or low counts lack $\Gamma$ measurements (see Section \ref{subsec:Xobs}) and are not included. The solid orange line shows a linear regression fit excluding GH~47 (tagged by name), with the 1$\sigma$ confidence level given by the yellow-shaded area. Our sample shows weak hints of a correlation ($p_{\rm null}=0.087$). We extend the figure leftward to illustrate how the best-fit regression reaches values of $\Gamma \sim 1.0$ near our X-ray-weakest targets' upper limits on $\daox$ ($\lesssim -0.7$), consistent with the X-ray stacking analysis performed by \citet{Plotkin16}. The dashed orange line shows an alternate version of the regression including GH~47.
\label{fig:Gamma_vs_daox}}
\end{figure}

Therefore, while we do not expect the existing X-ray data to offer a ``smoking gun," we still search for additional signs of X-ray shielding in our sample by comparing the observed X-ray spectrum and $\daox$. The shielding medium should modify the intrinsic spectrum associated with the corona, hardening (flattening) it as soft X-rays are more likely to be absorbed, scattered, and/or reflected away (see, e.g., \citealt{Gibson09}, particularly their Figure 7 and related discussion on X-ray absorption in BAL quasars). Some variance is expected in the relationship, as the extent to which the spectrum is modified will depend on the covering factor and column density of the obscuring medium. Figure \ref{fig:Gamma_vs_daox} examines the dependence of the observed X-ray photon index, $\Gamma$, on $\daox$. We caution that not only is there a higher degree of uncertainty for the X-ray weaker objects, but this plot also does not incorporate our X-ray weakest, non-detected objects. Additionally, we find one target (GH47, tagged by name in Figure \ref{fig:Gamma_vs_daox}) that may be an outlier: while this source does not appear to be extremely X-ray weak, the shape of its Chandra spectrum is relatively hard ($\Gamma=0.36$), which may imply a high level of absorption.\footnote{The Chandra observation of GH~47 has too few counts to quantify absorption directly from the data, but \citet{Ludlam15} examined this object using multi-epoch XMM-Newton observations and found it to have a 0.3--10~keV spectrum that may suggest an obscured, Type 2 AGN (see Appendix \ref{sec:app}).} 

Even given the above caveats, it appears plausible that the X-ray-weaker objects in our sample show generally harder (flatter) X-ray spectra. The solid orange line in Figure \ref{fig:Gamma_vs_daox} shows a linear fit via orthogonal distance regression (weighted by uncertainty), excluding GH~47. We find only a hint of a correlation at the $p_{\rm null}=0.087$ level from Kendall's Tau. However, \citet{Plotkin16} performed a photon stacking analysis of X-ray non-detected, primarily moderate- to high-Eddington \citetalias{GH07s} AGNs and found a flat, albeit highly uncertain, stacked spectrum ($\Gamma = 1.0^{+1.1}_{-0.5}$). Following the linear fit on Figure \ref{fig:Gamma_vs_daox} down to our X-ray-weakest, non-detected targets' upper limits on $\daox$ ($\lesssim -0.7$) leads to values of $\Gamma \sim 1.0$, consistent with the \citet{Plotkin16} analysis.\footnote{While this stacking analysis may conflict with our assumption of the weighted sample mean for our X-ray non-detected objects ($\Gamma=1.9$; see Section \ref{subsec:Xobs}), we have already included uncertainty on $\Gamma$ in our error analysis (see Section \ref{sec:results}) and found that it is unlikely to significantly alter our main result. Furthermore, while we could use the fit found here to refine our assumed $\Gamma$ for non-detections based on their $\daox$, the issue becomes circular as our $\daox$ calculations are already dependent on $\Gamma$.} While we caution against drawing strong conclusions from this assessment, we do consider it suggestive of the X-ray-weak objects having, on average, spectral characteristics consistent with X-ray obscuration. More sensitive observations are needed to better determine the sample's X-ray spectral properties.

Finally, we find that the available observations already imply a relative limit on the radial extent of a physical X-ray shielding region. \citet{Ludlam15} performed X-ray variability analysis of a subsample of \citetalias{GH07s} mBH AGNs (including three objects overlapping ours: GH~47, GH~211, and GH~213) and found short-term variability (timescales of $\sim$\,100--1000~s) suggesting X-ray emission on scales $\sim$\,0.2--2~au. The X-ray-obscuring medium must reside outside this range, yet the \citetalias{GH07s} sample was identified as optically unobscured from SDSS spectra. Furthermore, \citet{Laor08} (see Section 3.5.2 therein) estimate that even the smallest possible scale emission predicted at 5~GHz, e.g., from a compact radio corona or jet base, should originate from a region $\gtrsim 100$ times the extent of the X-ray emitting core. The tentative $\Rx$--$\daox$ anticorrelation itself disfavors shielding of nuclear radio emission. The putative shielding region therefore appears to be compact, affecting coronal X-ray emission but neither optical nor radio. Critically, however, no UV spectral observations exist for any significant portion of the \citetalias{GH07s} sample at the time of writing (to our knowledge). Obtaining such data will allow us to better constrain the spectral energy distribution of these AGNs and further eliminate suspicion of a more extended obscuring medium in the X-ray weak population.

\subsubsection{The Influence of X-ray Variability} \label{subsec:variability}

While our discussion to this point has assumed that our sample is not particularly variable in hard (2--10~keV) X-rays, \citet{Ludlam15} show indications of long-term (timescales of months--years) variability in soft (0.5--2~keV) X-rays for several \citetalias{GH07s} AGNs (including two of our targets, GH~211 and GH~213). As discussed in Section \ref{subsec:properties}, X-ray variability could have a significant effect on our uncertainty, moving data points diagonally in Figure \ref{fig:main_result}(b). To dominate our results, it must weaken our X-ray-weakest targets by up to (or possibly exceeding, in the case of non-detections) $\sim$\,100$\times$ the $\alpha_{\rm ox,qso}$ expected from their optical spectra (between their SDSS and Chandra observation epochs, on timescales of $\lesssim 10$ years). Such variability is not impossible, considering our discussion of NLS1 dynamic coronae in Section \ref{subsec:weakness} as well as recent analysis of a changing-look AGN in which the X-ray corona appeared to be destroyed and recreated on a time scale of $\sim$\,1~yr (\citealt{Ricci20}; see also \citealt{Ricci23} and references therein). However, we find that $\gtrsim$\,40$\%$ of our sample must be weakened to $10 \lesssim \fweak \lesssim 100$ in order to drive the observed $\log \Rx$--$\daox$ anticorrelation at a similar confidence level.\footnote{As determined by the following test: For all 18 objects, we scale X-ray fluxes by $\fweak$ to a value consistent with $\daox \approx 0$. We then re-run the Monte Carlo simulations described in Section \ref{subsec:properties} while restoring objects with $\fweak > 10$ one at a time to their observed X-ray flux levels, and we stop when a fit and confidence level consistent with our primary result (Table \ref{tab:fits}) is achieved. We find this is reached with $\approx$\,8 objects restored to X-ray weakness.} Follow-up observations of this sample will help place constraints on whether such an extreme level of X-ray variability in the mBH population is realistic.

We also note that it is currently not clear if there could be a relationship between X-ray and radio variability. If the X-ray and radio emission mechanisms are coupled but the radio occupies a larger physical scale, we would (perhaps na{\"i}vely) expect corresponding radio variability that is ``smeared out" by comparison; i.e., it will have smaller amplitude and longer delay/duration than that seen in X-ray. \citet{Behar15} examined highly X-ray-variable AGNs and found that the $\Rx \sim$ constant trend held at 95GHz (corresponding, per \citealt{Laor08}, to radio emission scales as small as $\sim$\,$10^{-4}$~pc, or $\sim$\,200~au), but noted a need for simultaneous radio and X-ray monitoring to better understand the link between coronal X-ray and radio emission. Such studies are being undertaken but have yet to produce a solution \citep[e.g.,][]{Behar20, Chen22, Petrucci23}.

\subsubsection{Slim-Disk Accretion and Comparison to Weak Emission-Line Quasars} \label{subsec:slim_disk}

\citet{Desroches09} and \citet{Plotkin16} suggest that a ``slim" accretion disk may be responsible for the observed X-ray properties of high-Eddington mBHs, for reasons we describe below. Accretion disks in the near- or super-Eddington regime ($\ledd \gtrsim 0.3$) are expected to become radiatively inefficient, thickening geometrically at small radii \citep[e.g.,][]{Abramowicz88, Czerny19} and launching wide-angle outflows \citep[e.g.,][]{Murray95, Castello-Mor17, Giustini19, Jiang19, Naddaf22}. The outer disk may remain geometrically thin, but the vertical extent at small radii can reach $H/R \approx 0.3$ \citep[e.g., Section 6 of][]{Abramowicz13}, where $H$ is the height of the disk from the equatorial plane and $R$ is its radial extent. Identifying such objects presents a challenge due to an observational degeneracy created by radiative inefficiency: in the slim-disk scenario, the true mass accretion rate will likely be higher (compared to a radiatively efficient thin disk) than the observed luminosity implies. That is, some slim-disk objects may only display a moderate Eddington ratio.

The slim-disk scenario was proposed by \citet{Luo15} as one possible explanation for weak emission-line quasars (WLQs), a population of  AGNs with preferentially weakened high-ionization UV broad line emission (such as Ly$\alpha$ and \ion{C}{4} $\lambda$1549; see, e.g., \citealt{McDowell95, Fan99, DS09}) and a substantial X-ray-weak fraction ($\sim$\,50\%; \citealt{Ni18, Ni22, Pu20}) with signs of X-ray obscuration \citep[e.g.,][]{Luo15}. In this scenario, the inflated inner region of the disk and the related outflows will shield the broad emission-line region (BELR) from ionizing extreme-UV/X-ray radiation and obscure X-ray emission from our line of sight at larger inclination angles \citep[e.g.,][]{Wu11, Wu12, Luo15, Ni18, Ni22}. 

\citet{Plotkin16} compared the X-ray properties of \citetalias{GH07s} mBHs to those of WLQs and noted a number of similarities, including: a wide spread in $\aox$ with a prominent X-ray-weak tail \citep[e.g.,][]{Wu11, Wu12, Luo15}; generally softer X-ray spectra ($\Gamma \gtrsim 1.5$) in the X-ray-normal WLQ subpopulation, while stacking of the X-ray-weak subpopulation suggests harder spectra ($\Gamma \lesssim 1.5$) from X-ray absorption and/or reflection \citep[e.g.,][]{Luo15, Ni22}; and prevalence of high Eddington ratios ($\ledd \gtrsim 0.3$; e.g., \citealt{Shemmer10, Luo15, Plotkin15, Marlar18}).\footnote{Indications that $\ledd$ in our sample may be underestimated by up to an order of magnitude are discussed in Section \ref{subsec:samp}.} While these comparisons are not conclusive, further opportunities abound. If the WLQ and \citetalias{GH07s} mBH AGN populations share these similarities due to common accretion states, they may have other, as-yet-unexplored observables in common. For example, if a slim disk is responsible for shielding the BELR and producing the eponymously weak UV broad-line emission in WLQs, a similar effect may be found in high-accretion-rate mBH AGNs. As with WLQs, we might expect to see weakened broad lines with no correlation between line weakness and observed X-ray weakness, suggesting the BELR receives less ionizing radiation even when we observe the AGN to be X-ray normal \citep[e.g.,][]{Luo15, Ni18, Paul22}. However, since a hotter disk peaking in soft X-ray is expected at lower BH mass \citep[e.g.,][]{Done12}, the ionizing continuum in mBHs may differ from that of more massive quasars \citep[e.g.,][]{Wu24}. Therefore, to better characterize the ionization state of the mBH BELR, observations of the UV continuum and high-ionization broad-line emission (e.g., Ly$\alpha$ and/or \ion{C}{4}) are highly desirable to complement the low-ionization line emission (e.g., H$\beta$) already available from SDSS spectra.

Finally, we return briefly within this context to the topic of variability. The origin of X-ray variability in AGNs has yet to be fully understood and may depend on a number of physical effects. While intrinsically X-ray-weak objects may experience X-ray variability due to changes in disk-corona coupling, some recent studies indicate WLQs (or otherwise high-Eddington quasars) can experience extreme, rapid X-ray variability that seems to be linked to wide-angle, central-engine outflows such as disk winds \citep[e.g.,][]{Miniutti12, Liu22, Mao22, Vito22, Wang22, Giustini23, Huang23, Zhang23} and/or slight variations in the vertical thickness of the inner accretion disk \citep[e.g.,][]{Liu19, Ni20}. If our results do signify slim-disk accretion in \citetalias{GH07s} mBH AGNs, there may therefore be an increased likelihood of outflows contributing to X-ray weakness and/or variability. Conversely, \citet{Laurenti22} examined a sample of super-Eddington ($\ledd > 1$), supermassive ($M_{\rm BH} > 10^{8}~M_\odot$) AGNs at $0.4 \leq z \leq 0.75$ and found that $\sim$\,30$\%$ of their sample was X-ray weak ($\fweak > 10$) with steep X-ray spectra, implying intrinsic X-ray weakness. They suggested a scenario in which strong disk (or failed) winds related to super-Eddington accretion were directly responsible for an intrinsically-weak corona by reducing the number of seed photons available for Compton upscattering \citep[see also][]{Nardini19, Zappacosta20}.

Still, the ability of lower-mass AGNs such as these to launch disk winds is not yet established; their accretion disks could be too hot and low-luminosity to drive strong winds even at high accretion rates \citep[e.g.,][]{Giustini19, Naddaf22}. We again stress the importance of obtaining UV spectral observations of the \citetalias{GH07s} mBH AGN sample, as the presence or absence of asymmetries in \ion{C}{4} (which is often seen to be heavily blueshifted in wind-dominated WLQs; e.g., \citealt{Richards11, Plotkin15, Rivera22, Matthews23}) will help us investigate the existence of outflows. A discovery of the absence of winds in an mBH slim-disk scenario would also lend support to super-Eddington accretion as a viable channel for SMBH growth, as the un-radiated accretion power stands a better chance of being advected directly into the black hole rather than being driven away in an outflow \citep[e.g.,][]{Madau14}.

\section{Summary and Conclusions} \label{sec:summ}

We performed a multiwavelength pilot study of the curious X-ray-weak ``tail" observed in populations of low-mass AGNs. For a sample of high-Eddington-ratio, high-confidence mBHs ($M_{\rm BH} \sim 10^{6}~M_\odot$) from the catalog of \citet{GH07s}, we combined new and archival VLA radio data with archival Chandra X-ray data covering the full range of observed X-ray weaknesses (as described by the quantity $\daox$). Existing empirical correlations indicate that, for SMBHs, the radio--X-ray luminosity ratio ($\Rx = \lrlx$) is approximately constant at a given radio emission scale in radio-quiet, high-Eddington AGNs (see Section \ref{sec:intro}), and we have examined our sample of mBHs in that context. We summarize our findings as follows:

\begin{itemize}
\setlength{\itemsep}{0pt}
    \item  Most of our targets show unambiguous compact radio emission consistent with a point source and radio--optical properties consistent with radio-quiet, high-Eddington-ratio AGNs (Section \ref{sec:results}).
    \item We find a tentative anticorrelation ($p_{\rm null} \approx 0.008$) between $\Rx$ and $\daox$ in our sample (Section \ref{subsec:properties}). This observation suggests that X-ray emission is being preferentially weakened compared to radio emission (Section \ref{sec:discussion}).
\end{itemize}

Most of the X-ray-weak, $\Rx$-intermediate objects in our sample are still clearly radio-quiet relative to the optical, and we find no indication of the observed anticorrelation being directly influenced by star formation, relativistic beaming, or differences in core radio emission mechanism (Section \ref{subsec:mechanism}). 

We examine two primary scenarios to explain our results: intrinsic X-ray weakness and X-ray shielding. The $\Rx$--$\daox$ relationship we observe does not appear to be consistent with an intrinsic X-ray weakness scenario (Section \ref{subsec:weakness}), although we cannot exclude the possibility that there are some intrinsically weak individual sources. We also argue that variability is unlikely to be directly responsible (Section \ref{subsec:variability}). Instead, we argue that on average our sample appears consistent with a simple prediction model for X-ray shielding (see Eq.~\ref{eq5}), suggesting the ``slim disk" scenario in which X-rays are preferentially obscured from our line of sight by the inner radii of the accretion disk. We extend the discussion of \citet{Plotkin16} comparing high-Eddington mBH AGNs to WLQs and speculate that the two populations may share other characteristics that have yet to be explored in mBH AGNs (e.g., preferentially weak high-ionization broad UV emission lines; see Section \ref{subsec:slim_disk}). 

We conclude that the tentative $\Rx$--$\daox$ anticorrelation suggests a new observational signature for finding high-accretion-rate AGN populations via radio--X-ray analysis. This exploratory result justifies gathering larger samples of radio-quiet AGNs with significant X-ray-weak percentages, including NLS1s, quasars, and WLQs, to perform comparable tests.

\hfill \break
\indent We are extremely grateful to the anonymous referee for thorough and constructive comments that helped improve this manuscript. J.D.P. thanks the NRAO staff for hosting the 9th VLA Data Reduction Workshop. J.D.P. and R.M.P. acknowledge support from the National Science Foundation under grant number 2206123. W.N.B. acknowledges support from Chandra X-ray Center grant GO2-23083X and National Science Foundation grant AST-2106990. L.C.H. was supported by the National Science Foundation of China (11991052, 12011540375, 12233001), the National Key R\&D Program of China (2022YFF0503401), and the China Manned Space Project (CMS-CSST-2021-A04, CMS-CSST-2021-A06). The scientific results reported in this paper are based in part on data obtained from the Chandra Data Archive, using software provided by the Chandra X-ray Center (CXC) in the application package CIAO. The list of Chandra datasets employed is available in Chandra Data Collection (CDC) 254 at\dataset[doi:10.25574/cdc.254]{https://doi.org/10.25574/cdc.254}. The NRAO is a facility of the National Science Foundation operated under cooperative agreement by Associated Universities, Inc. Funding for the SDSS has been provided by the Alfred P. Sloan Foundation, the Participating Institutions, the National Science Foundation, the U.S. Department of Energy, the National Aeronautics and Space Administration, the Japanese Monbukagakusho, the Max Planck Society, and the Higher Education Funding Council for England. The SDSS web site is \url{www.sdss.org}. For this research, we made use of the following resources: the SIMBAD database and the VizieR catalogue access tool, both operated at CDS, Strasbourg, France; the NASA/IPAC Extragalactic Database (NED), which is funded by the National Aeronautics and Space Administration and operated by the California Institute of Technology; the \texttt{Python} language along with \texttt{Astropy} \citep{astropy13, astropy18}, \texttt{NumPy} \citep{Numpy}, and \texttt{SciPy} \citep{scipy}; \texttt{SAOImageDS9} \citep{saoimage_ds9}; and \texttt{TOPCAT} \citep{topcat}.

\appendix
\restartappendixnumbering

\section{Notes on Individual Objects} \label{sec:app}

\begin{figure*}[p!]
\epsscale{1.17}
\plotone{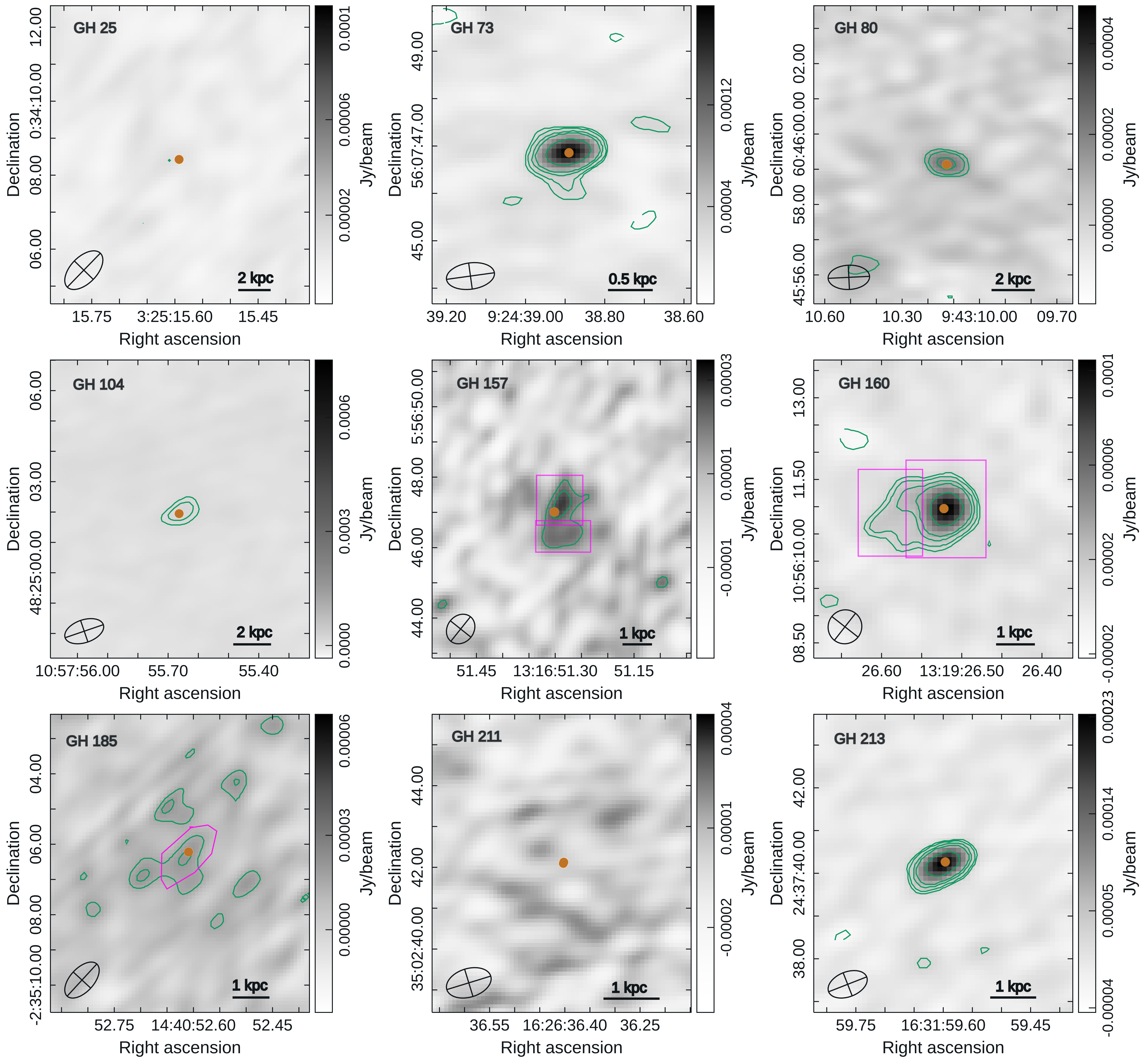}
\caption{Radio images for the first 9 new observations (object order follows the earlier data tables). North is up and east is left. For each image, GH ID is given in the upper left corner, synthesized beam size is shown in the lower left corner, and estimated projected physical scale at the target's location is illustrated in the lower right corner. Green contours are applied at -3 (dashed), 3, 4, 6, 8, 16, and 32$\times$ the image rms ($\sigma_{\rm rms}$). The orange dot indicates galaxy center coordinates from SDSS. The greyscale tone map indicates flux density in Jy~bm$^{-1}$.
\textbf{Top left:} GH~25, $\sigma_{\rm rms}$ = 0.004 mJy, beam size $1\farcs28 \times 0\farcs72$. \textbf{Top middle:} GH~73, $\sigma_{\rm rms}$ = 0.007 mJy, beam size $1\farcs02 \times 0\farcs56$. \textbf{Top right:} GH~80, $\sigma_{\rm rms}$ = 0.004 mJy, beam size $1\farcs18 \times 0\farcs69$. 
\textbf{Mid left:} GH~104, $\sigma_{\rm rms}$ = 0.004 mJy, beam size $1\farcs06 \times 0\farcs59$. \textbf{Center:} GH~157, $\sigma_{\rm rms}$ = 0.006 mJy, beam size $0\farcs89 \times 0\farcs75$. \textbf{Mid right:} GH~160, $\sigma_{\rm rms}$ = 0.004 mJy, beam size $0\farcs63 \times 0\farcs61$. 
\textbf{Bottom left:} GH~185, $\sigma_{\rm rms}$ = 0.005 mJy, beam size $1\farcs24 \times 0\farcs68$. \textbf{Bottom middle:} GH~211, $\sigma_{\rm rms}$ = 0.008 mJy, beam size $1\farcs13 \times 0\farcs70$. \textbf{Bottom right:} GH~213, $\sigma_{\rm rms}$ = 0.009 mJy, beam size $0\farcs96 \times 0\farcs58$.
\label{fig:radio_images1}}
\end{figure*}

\begin{figure*}[p!]
\epsscale{1.17}
\plotone{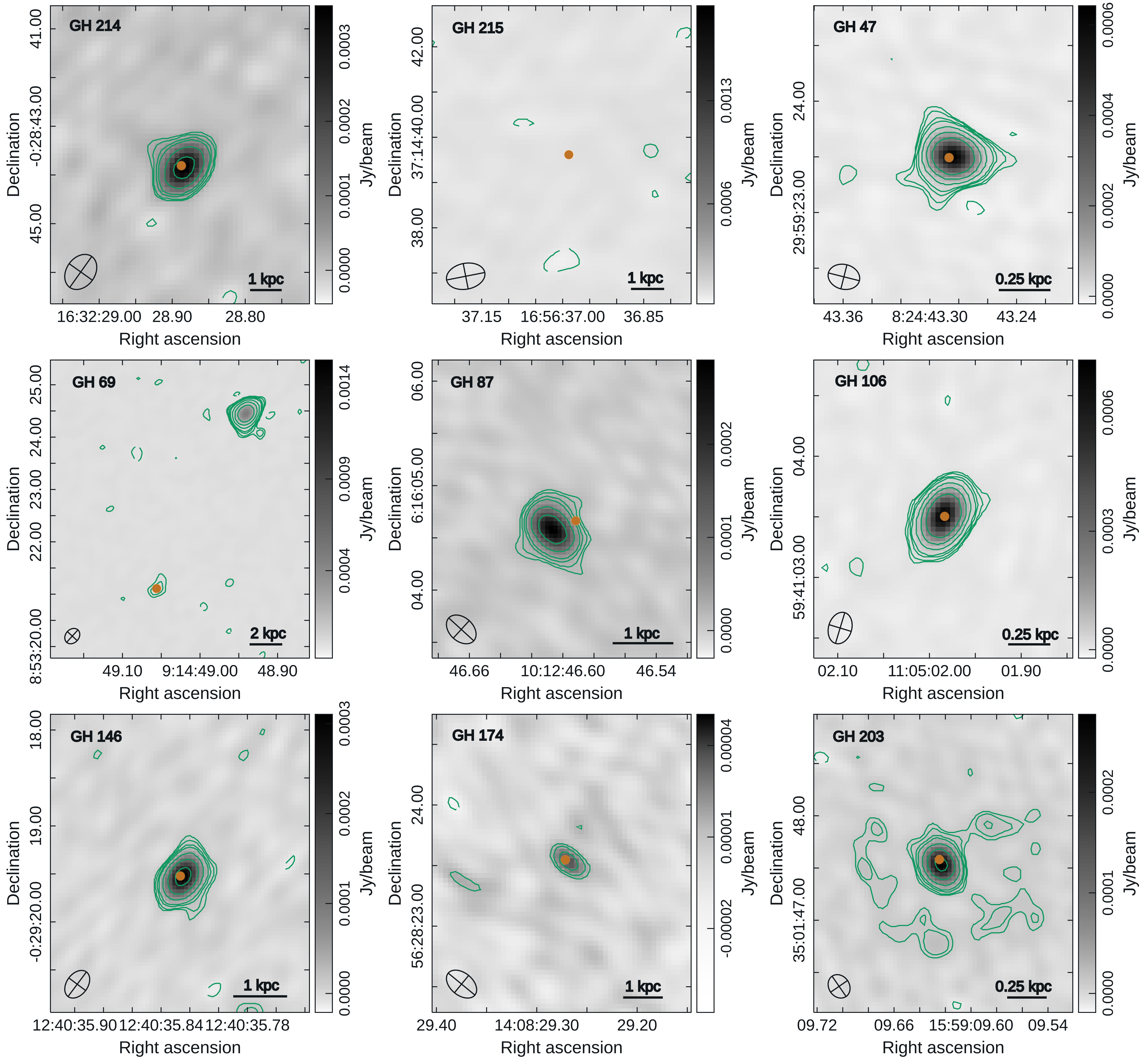}
\caption{Radio images for the remaining 2 new observations and all archival observations (object order follows the earlier data tables). North is up and east is left. For each image, GH ID is given in the upper left corner, synthesized beam size is shown in the lower left corner, and estimated projected physical scale at the target's location is illustrated in the lower right corner. Green contours are applied at -3 (dashed), 3, 4, 6, 8, 16, 32, and 64$\times$ the image rms ($\sigma_{\rm rms}$). The orange dot indicates galaxy center coordinates from SDSS. The greyscale tone map indicates flux density in Jy~bm$^{-1}$.
\textbf{Top left:} GH~214, $\sigma_{\rm rms}$ = 0.009 mJy, beam size $0\farcs78 \times 0\farcs59$. \textbf{Top middle:} GH~215, $\sigma_{\rm rms}$ = 0.010 mJy, beam size $0\farcs86 \times 0\farcs57$. \textbf{Top right:} GH~47, $\sigma_{\rm rms}$ = 0.003 mJy, beam size $0\farcs29 \times 0\farcs22$. 
\textbf{Mid left:} GH~69, $\sigma_{\rm rms}$ = 0.004 mJy, beam size $0\farcs32 \times 0\farcs26$. \textbf{Center:} GH~87, $\sigma_{\rm rms}$ = 0.005 mJy, beam size $0\farcs32 \times 0\farcs23$. \textbf{Mid right:} GH~106, $\sigma_{\rm rms}$ = 0.003 mJy, beam size $0\farcs27 \times 0\farcs19$. 
\textbf{Bottom left:} GH~146, $\sigma_{\rm rms}$ = 0.004 mJy, beam size = $0\farcs32 \times 0\farcs22$. \textbf{Bottom middle:} GH~174, $\sigma_{\rm rms}$ = 0.005 mJy, beam size $0\farcs29 \times 0\farcs18$. \textbf{Bottom right:} GH~203, $\sigma_{\rm rms}$ = 0.004 mJy, beam size $0\farcs24 \times 0\farcs2$.
\label{fig:radio_images2}}
\end{figure*}

Radio images of our 11 new and 7 archival VLA observations are shown in Figures \ref{fig:radio_images1} and \ref{fig:radio_images2} (consistent with our data tables, we give the new observations first, followed by the archival). Below, we discuss a handful of individual sources that merit further comment.

\subsection{New VLA Observations} \label{subsec:app_new}

\subsubsection{GH 157} \label{subsec:gh157}
At 10~GHz, GH~157 shows a primary component consistent with a resolved source at the SDSS coordinates, as well as a faint secondary emission component that may be consistent with a jet or jet remnant morphology. We utilized two Gaussians to fit this object, illustrated by the purple region boxes in Figure \ref{fig:radio_images1}(center). The secondary component has a peak flux density that is only marginally detected ($\approx$\,3$\sigma_{\rm rms}$). The radio properties reported in Table \ref{tab:obslog} and throughout the paper are based on measurement of the primary component. 

\subsubsection{GH 160} \label{subsec:gh160}
At 10~GHz, GH~160 shows a primary component consistent with an unresolved point source at the SDSS coordinates, as well as a faint secondary emission component that may be consistent with a jet or jet remnant morphology. We utilized two Gaussians to fit this object, illustrated by the purple region boxes in Figure \ref{fig:radio_images1}(mid-right). The radio properties reported in Table \ref{tab:obslog} and throughout the paper are based on measurement of the primary component. This object was originally presented as an X-ray non-detection by \citet{Dong12a}, but we have redetermined it as a detection based on our analysis (as described in Section \ref{subsec:Xobs}). 

\subsubsection{GH 185} \label{subsec:gh185}
With weighting set to \texttt{robust=1.5} during imaging, GH~185 shows diffuse extended emission at 10~GHz with no obvious core emission component. For our measurements, we have selected the component most resembling a point-like source nearest the SDSS coordinates, as illustrated by the purple selection region in Figure \ref{fig:radio_images1}(bottom-left). We also experimented with the application of a 2D Gaussian taper in the uv-domain; the resulting image clearly showed a resolved emission source with increased sensitivity but reduced resolution (synthesized beam size increased by $\sim$\,2$\times$). From this test image we obtained a peak flux density $S_{\rm p} = 0.048 \pm 0.007$~mJy using CASA task \texttt{imstat}, but given the appearance of the original non-tapered image, we are unable to confidently attribute this measurement to the AGN core location/activity. We also note that this object was not detected in X-ray by Chandra. It is apparent that more sensitive observations are necessary in both wavebands to better understand the nature of GH~185's emission and verify AGN activity.

\subsubsection{GH 211} \label{subsec:gh211}
GH~211 has a ROentgen SATellite (ROSAT) X-ray detection and four XMM-Newton observations in three different years (2007, 2011, 2012) in addition to the Chandra observation. \citet{Ludlam15} noted strong long-term X-ray variability (factor of nearly 10 in the 0.5--2~keV band) between the four XMM-Newton observations. Per its narrow optical emission lines, this object is a composite galaxy on the BPT diagram, and it is not detected at 10~GHz ($S_{\rm p} < 0.025$~mJy at the 3$\sigma_{\rm rms}$ level).

\subsubsection{GH 213} \label{subsec:gh213}
GH~213 was originally presented as an X-ray non-detection by \citet{Dong12a}, but we have redetermined it as a detection based on our own X-ray analysis (as described in Section \ref{subsec:Xobs}). This object has a ROSAT detection and was also observed in 2011 by XMM-Newton. \citet{Ludlam15} note that it displays a factor of $\sim$\,4 long-term variability in the 0.5--2~keV band between the ROSAT and XMM-Newton observations. At 10~GHz, we find a simple point source.

\subsubsection{GH 214} \label{subsec:gh214}
GH~214 was originally presented as an X-ray detection by \citet{Dong12a}, but we have redetermined it as a non-detection based on our own analysis. At 10~GHz, we find a marginally resolved object consistent with a point source (peak-to-integrated flux ratio 0.95). It is $\Ro$-intermediate ($\Ro \approx 35$), and we tag it by name for easy identification in Figures \ref{fig:main_result} and \ref{fig:Radio_loud}(b).

\subsection{Archival VLA Observations} \label{subsec:app_archv}

\subsubsection{GH 47} \label{subsec:gh47}
GH~47 is marginally resolved at 9~GHz (peak-to-integrated flux ratio 0.83), consistent with a point source at the SDSS coordinates. We find it to be borderline $\Ro$-intermediate ($\Ro = 10.3$). \citet{Ludlam15} examined this object in X-ray using multi-epoch XMM-Newton observations and found it to have low variability and a 0.3--10~keV spectrum that may suggest an obscured, Type 2 AGN (see their Figure 2 and related discussion). We therefore exclude it as an outlier from our $\Gamma$--$\daox$ analysis in Section \ref{subsec:weakness}. Additionally, its exclusion from our $\Rx$--$\daox$ analysis is tested (along with three other $\Ro$-intermediate objects) in Section \ref{subsec:mechanism}.

\subsubsection{GH 69} \label{subsec:gh69}

\begin{figure}[t]
\epsscale{1.15}
\plotone{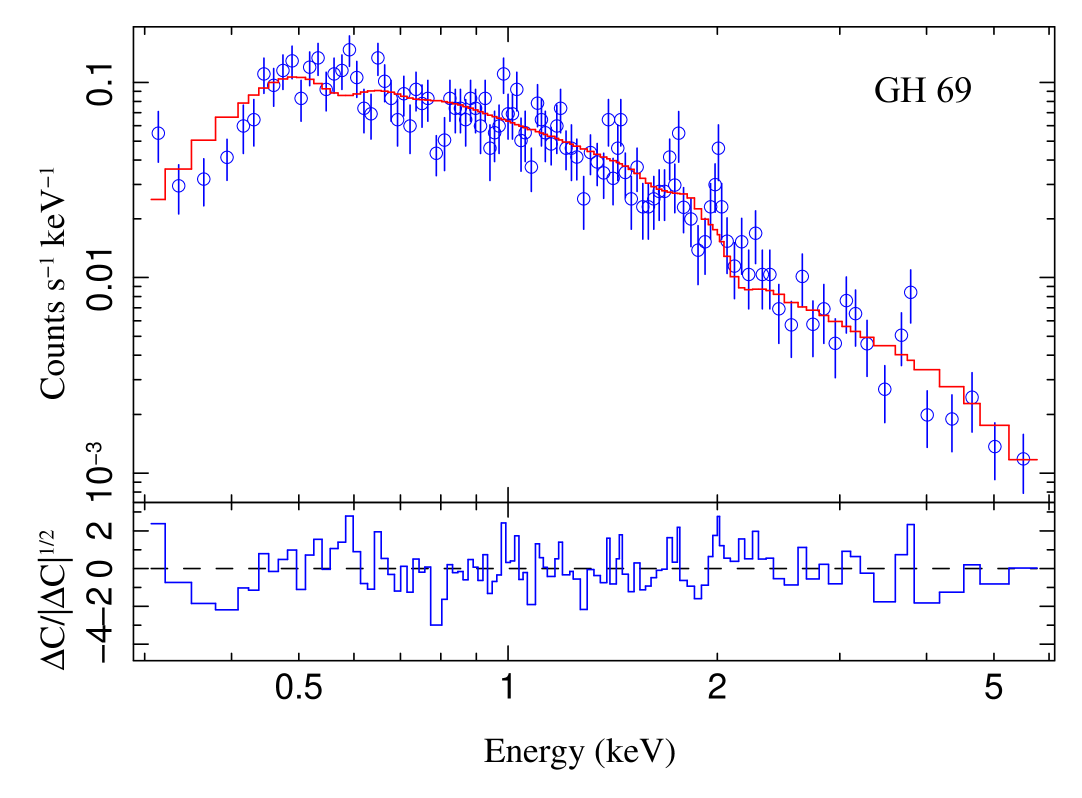}
\caption{Spectral fit for the GH~69 Chandra observation (see Appendix \ref{subsec:gh69}). Residuals are displayed as $\Delta C / |\Delta C|^{1/2}$, where $C$ is the Cash statistic.
\label{fig:gh69}}
\end{figure}

At 9~GHz, we find a faint ($\sim$\,4$\sigma_{\rm rms}$) detection at the target's SDSS coordinates, marked by the orange dot in the south of Figure \ref{fig:radio_images2}(mid-left). However, we note that prior to cleaning this image, the target region was partly obscured by a sidelobe (from the convolved point-spread-function, a.k.a. ``dirty beam") of the brighter source seen in the northwest of the image. While it appears that the sidelobe has been effectively removed, we caution that the cleaning process may have impacted the flux observed at the target location. We also note that the coordinates of this brighter source coincide with a different galaxy, SDSS J091448.94+085324.4. In X-ray, the addition of a multi-temperature accretion disk component was deemed sufficient for this object's spectral fit ({\tt tbabs*ztbabs*[diskbb+powerlaw]}; see Figure \ref{fig:gh69}), and no further or more complicated models were pursued. 

\subsubsection{GH 87} \label{subsec:gh87}
The phase calibrator for GH~87's VLA observation shows two point sources, which precludes automatic calibration through the CASA pipeline. We performed self-calibration on the phase calibrator to create a suitable two-point model, then applied this model when manually calibrating the target visibilities. We find 9~GHz emission consistent with a point source.

\subsubsection{GH 203} \label{subsec:gh203}
GH~203 (a.k.a. Mrk~493) is a relatively well-studied object. To our knowledge, however, the VLA X-band image has not previously been published. We find a marginally resolved (peak-to-integrated flux ratio 0.75) central primary emission component, surrounded by faint extended emission that may coincide with optical--UV structure visible in Hubble Space Telescope imagery (see \citealt{Munoz07}). \citet{Berton18} examined this object across 1.4--5~GHz radio bands; they show that it also appears extended at 5~GHz (see their Figure C.33, left panel). For the core component, we find a steep 9~GHz integrated-flux spectral slope ($\alpha = -1.13 \pm 0.24$) that is consistent with both their in-band (5~GHz) and broadband (1.4--5~GHz) integrated-flux slopes. On the other hand, their 5~GHz peak-flux slope is steeper than our 9~GHz peak-flux slope, implying that the core radio spectrum flattens from 5 to 9~GHz (which we might expect if higher frequencies originate from more compact scales).

\bibliography{mainref.bib}{}
\bibliographystyle{aasjournal}

\end{document}